\documentclass[12pt]{article}
\usepackage{amssymb}
\usepackage{amsmath}
\usepackage{cite}
\usepackage{tensor}
\usepackage{hyperref}
\usepackage{amsfonts}
\usepackage{soul}
\usepackage{tikz-cd}

\newcommand{\Z}{\mathbb{Z}}
\newcommand{\klein}{\mathbb{Z}_2\times\mathbb{Z}_2}
\newcommand{\kerd}{\text{ker}(\delta)}
\newcommand{\cokerd}{\text{coker}(\delta)}
\newcommand{\aut}{\text{Aut}}
\newcommand{\xmod}{\Gamma=(\Gamma_1\xrightarrow{\delta} \Gamma_0)}
\input xy
\xyoption{all}
\usepackage{enumitem}

\def\sqr#1#2{{\vcenter{\vbox{\hrule height.#2pt
            \hbox{\vrule width.#2pt height#1pt \kern#1pt
                  \vrule width.#2pt}\hrule height.#2pt}}}}

\def\sqra#1#2#3{{\vcenter{\vbox{\hrule height.#2pt
            \hbox{\vrule width.#2pt height#1pt \kern5pt %\kern#1pt
#3
                  \vrule width.#2pt}\hrule height.#2pt}}}}

\numberwithin{equation}{section}

\numberwithin{table}{section}

\begin{document}

\begin{center}

{\large\bf Three-dimensional orbifolds by 2-groups}

\vspace*{0.2in}

Alonso Perez-Lona, Eric Sharpe

\begin{tabular}{l}
Department of Physics, MC 0435\\
850 West Campus Drive\\
Virginia Tech\\
Blacksburg, VA  24061
\end{tabular}

{\tt aperezl@vt.edu}, {\tt ersharpe@vt.edu}

$\,$

\end{center}

In this paper we generalize previous work on decomposition in three-dimensional orbifolds by 2-groups realized as analogues of central extensions, to orbifolds by more general 2-groups.  We describe the computation of such orbifolds in physics, state a version of the decomposition conjecture, and then compute in numerous examples, checking that decomposition works as advertised.

\begin{flushleft}
March 2023
\end{flushleft}

\newpage

\tableofcontents

\newpage

\section{Introduction}

This paper concerns decomposition, the observation that a local $d$-dimensional quantum field theory
with a global $(d-1)$-form symmetry is equivalent to (``decomposes into'') a disjoint union of other local quantum field theories, known as universes.  Now, any disjoint union of quantum field theories has a global $(d-1)$-form symmetry: the projection operators onto the different universes form topological dimension-zero operators corresponding to the $(d-1)$-form symmetry.  What makes decomposition particularly interesting is that this can happen in a local quantum field theory:  the path integral (on a connected spacetime) of a disjoint union of QFTs is a sum of path integrals, which one does not expect to be able to write as a single
path integral with a single action.  

Decomposition was first described in \cite{Hellerman:2006zs}, where it arose as part of efforts 
\cite{Pantev:2005rh,Pantev:2005wj,Pantev:2005zs} 
to resolve apparent physical inconsistencies in
string propagation on certain stacks known as gerbes, which encode higher-form symmetries geometrically (as the fibers of a fiber bundle), and which are realized physically as two-dimensional gauge theories and orbifolds in which a subgroup of the gauge group acts trivially.  
Since that time, decomposition has been further explored in a wide variety of theories in a variety of dimensions, and it has been applied to, for example, Gromov-Witten theory, gauged linear sigma models, elliptic genera, and anomaly resolution in orbifolds.  See for example the recent review 
\cite{Sharpe:2022ene} for references and further details.

The purpose of this paper is to study three-dimensional\footnote{
The quantum field theory is defined over a three-dimensional spacetime, as an orbifold of a low-energy effective sigma model, with a target space of any dimension, not necessarily three.
} orbifolds by 2-groups, generalizing results in \cite{Pantev:2022kpl} for orbifolds by special 2-groups
(specifically, analogues of central extensions).  These orbifolds gauge a combination of both an ordinary group as well as a `group' of one-form symmetries ('gerby' symmetries, in older terminology).  Further, we take the one-form symmetries to act\footnote{
In the language of defects, whether a one-form symmetry acts trivially or not in a three-dimensional theory is determined by its action on lines, 
see e.g.~\cite{Pantev:2022pbf} for a detailed discussion.
} trivially, so gauging them results in a theory with a global 2-form symmetry, and hence, in three dimensions, a decomposition.

Amongst other things, we will find a more general pattern of three-dimensional decompositions than
was observed in \cite{Pantev:2022kpl}.  There, in a given orbifold, the universes were all of the form
$[X/G]_{\omega}$, with the same orbifold group $G$, differing (at most) by choices of discrete torsion $\omega$.  Here, in our analysis of more general cases, we find examples in which the universes are different orbifolds, with different orbifold groups.

We begin in section~\ref{sect:basic-defn} by defining three-dimensional orbifolds by 2-groups, describing, for example, how to compute partition functions explicitly.  In section~\ref{sect:decomp-conj} we state a
conjecture for decomposition in these orbifolds, namely that for a 2-group $\Gamma$,
\begin{equation}
    {\rm QFT}\left( [X/\Gamma] \right) \: = \: 
    {\rm QFT}\left( \left[ \frac{ X \times \hat{K} }{G} \right]_{\omega} \right),
\end{equation}
where $G$, $K$ are determined by $\Gamma$, and $\omega$ indicates
three-dimensional discrete torsion (also determined ultimately by $\Gamma$).
In section~\ref{sec:examples} we verify that conjecture in a wide variety of examples, beginning with the cases in which the 2-group is equivalent to an ordinary group and,
separately, a centrally-extended group, checking that our methods duplicate (and generalize) existing results.  We then describe a variety of further examples, uncovering a more general pattern of decompositions than previously discussed in three-dimensional examples in \cite{Pantev:2022kpl,Pantev:2022pbf}.
Finally, in appendix~\ref{app:background} we provide some background material on 2-groups, to assist the reader with some of the more technical aspects of these computations.

Aside from a few general observations, in most of this paper we restrict to
2-group orbifolds $[X/\Gamma]$ without three-dimensional analogues of discrete torsion.  Generalizations are left for future work.

In passing, we should point out some pertinent topological field theories.
Dijkgraaf-Witten theory is an orbifold of a point, so its two-dimensional versions decompose, since they involve
a completely trivially-acting group, and hence a global one-form symmetry.  In particular, they completely  decompose, to a disjoint union of invertible field theories, a special case of a more general decomposition phenomenon in unitary two-dimensional topological field theories with semisimple local operator algebras
\cite{Durhuus:1993cq,Moore:2006dw,Komargodski:2020mxz,Huang:2021zvu}.
Similarly, three-dimensional 2-group orbifolds of a point also define topological field theories, and are versions of the Yetter model \cite{yetter,porter1,porter2}.  Since they involve trivially-acting gauged one-form symmetries, they have a global two-form symmetry, and so, as three-dimensional theories, they also decompose.  We will not belabor this point, but we will occasionally make use of results from Yetter models in our construction in this paper.

\section{Description of 3d orbifolds by 2-groups}  \label{sect:basic-defn}

\subsection{Basics}

In this section, we will define and describe three-dimensional orbifolds by 2-groups,
generalizing the work of \cite{Pantev:2022kpl} which studied special cases of 2-groups,
analogues of central extensions.
More details on 2-groups are provided in appendix~\ref{app:background};
see also e.g.~\cite{Sharpe:2015mja} and references therein.

Briefly, a 2-group is a generalization of an ordinary group, in which the group axioms are weakened to be isomorphisms rather than equalities.  A 2-group $\Gamma$ arises from a group $G$ and an abelian group $K$ along with $\alpha: G \rightarrow{\rm Aut}(K)$ and $\beta_{\Gamma} \in H^3(G,K)$ (with action on the coefficients determined by $\alpha$).
The 2-group $\Gamma$ can also be described as an
extension
\begin{equation}
    1 \: \longrightarrow \: BK \: \longrightarrow \: \Gamma \: \longrightarrow \: G \: \longrightarrow \: 1,
\end{equation}
classified by $\beta_{\Gamma}$.
The previous paper \cite{Pantev:2022kpl} specialized to
2-groups for which $\alpha$ was trivial; this paper considers more general cases.
That said, both this paper as well as \cite{Pantev:2022kpl} consider only 2-groups for which both $G$ and $K$ are finite, as both papers are devoted to orbifolds (instead of more general gauge theories).

In terms of notation, for $k \in K$, it is conventional to use
the notation
\begin{equation}
    {}^g k \: = \: \alpha(g)(k)
\end{equation}
for the action of $\alpha(g)$ on $k \in K$, for $g \in G$.

The group $G$ describes a $0$-form symmetry, whereas the abelian group $K$ describes a $1$-form symmetry. These symmetries are represented by codimension $1$ and $2$ defects, respectively. In terms of simple objects, in three dimensional QFT's, which are the focus of this paper, this corresponds to having surfaces labeled by elements $g\in G$, and lines characterized by group elements $a\in K$:
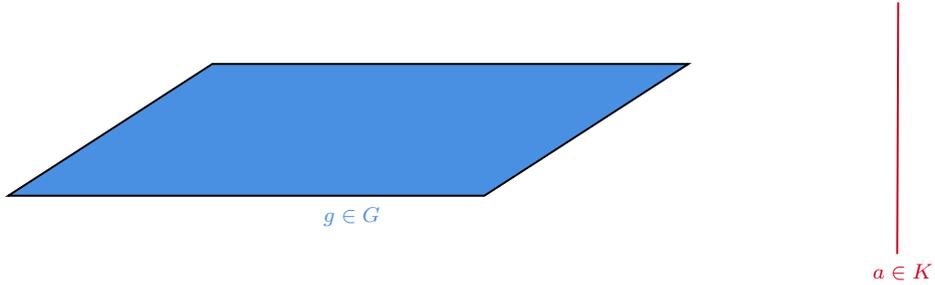
\begin{center}

\tikzset{every picture/.style={line width=0.75pt}} %set default line width to 0.75pt        

\begin{tikzpicture}[x=0.75pt,y=0.75pt,yscale=-1,xscale=1]
%uncomment if require: \path (0,279); %set diagram left start at 0, and has height of 279

%Shape: Parallelogram [id:dp9742748517354258] 
\draw  [fill={rgb, 255:red, 74; green, 144; blue, 226 }  ,fill opacity=1 ] (124.4,115) -- (364.5,115) -- (261.6,181.5) -- (21.5,181.5) -- cycle ;
%Straight Lines [id:da11758499736394601] 
\draw [color={rgb, 255:red, 208; green, 2; blue, 27 }  ,draw opacity=1 ]   (470.33,83.92) -- (469.83,210.92) ;

% Text Node
\draw (456.08,214.73) node [anchor=north west][inner sep=0.75pt]  [font=\scriptsize,color={rgb, 255:red, 208; green, 2; blue, 27 }  ,opacity=1 ]  {$a\in K$};
% Text Node
\draw (179,185.9) node [anchor=north west][inner sep=0.75pt]  [font=\scriptsize,color={rgb, 255:red, 208; green, 2; blue, 27 }  ,opacity=1 ]  {$\textcolor[rgb]{0.29,0.56,0.89}{g\in G}$};

\end{tikzpicture}

\end{center}

The remaining information $(\alpha,\beta_{\Gamma})$ describes mutual twists of such $0$- and $1$-form symmetries that allow considering $2$-group symmetries beyond the product $G\times BK$. The action $\alpha$ is pictorially represented via a \textit{piercing action}, which is visualized as an incoming line defect $a\in K$ traversing a surface defect $g\in G$ such that the outgoing line defect is the result of acting on $a$ by $g$:
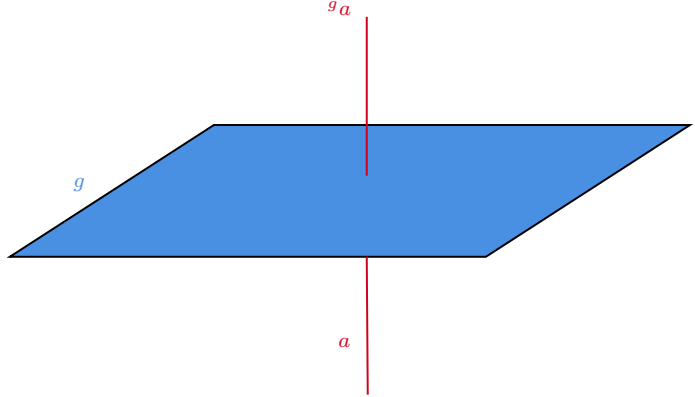
\begin{center}

\tikzset{every picture/.style={line width=0.75pt}} %set default line width to 0.75pt        

\begin{tikzpicture}[x=0.75pt,y=0.75pt,yscale=-1,xscale=1]
%uncomment if require: \path (0,279); %set diagram left start at 0, and has height of 279

%Shape: Parallelogram [id:dp9742748517354258] 
\draw  [fill={rgb, 255:red, 74; green, 144; blue, 226 }  ,fill opacity=1 ] (282.9,115) -- (523,115) -- (420.1,181.5) -- (180,181.5) -- cycle ;
%Straight Lines [id:da01797126840980323] 
\draw [color={rgb, 255:red, 208; green, 2; blue, 27 }  ,draw opacity=1 ]   (360,60.39) -- (360,140.57) ;
%Straight Lines [id:da894642080227321] 
\draw [color={rgb, 255:red, 208; green, 2; blue, 27 }  ,draw opacity=1 ]   (360,181.43) -- (360.5,251.08) ;

% Text Node
\draw (338.67,50.9) node [anchor=north west][inner sep=0.75pt]  [font=\scriptsize]  {$\textcolor[rgb]{0.82,0.01,0.11}{^{g} a}$};
% Text Node
\draw (344.17,220.9) node [anchor=north west][inner sep=0.75pt]  [font=\scriptsize]  {$\textcolor[rgb]{0.82,0.01,0.11}{a}$};
% Text Node
\draw (210.67,140.4) node [anchor=north west][inner sep=0.75pt]  [font=\scriptsize]  {$\textcolor[rgb]{0.29,0.56,0.89}{g}$};

\end{tikzpicture}

\end{center}

The extension class $\beta_{\Gamma}:G\times G\times G\to K$ describes the nontriviality of recombining the order of junctions, commonly known as $F$-moves. If such a class is trivial, any three defects $g,h,k\in G$ may be combined into a single defect $ghk\in G$ without any consideration of the order of association. If, on the other hand, the class is nontrivial, the different orders of associating such defects differ by a line defect $\beta(g,h,k)\in K$. A cross-section of such a situation is shown below, with the understanding that all defects extend in an additional unshown dimension (so that, as above, the blue defects are surfaces and the red defect is a line).
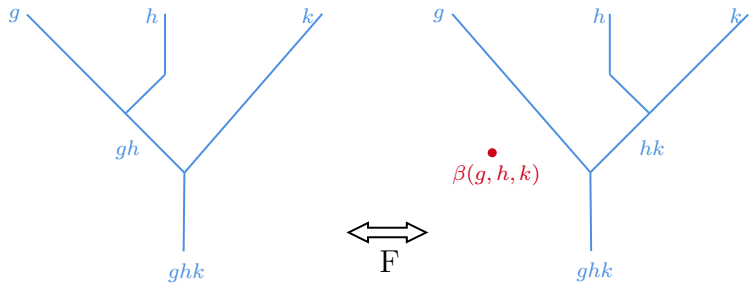
\begin{center}

\tikzset{every picture/.style={line width=0.75pt}} %set default line width to 0.75pt        

\begin{tikzpicture}[x=0.75pt,y=0.75pt,yscale=-1,xscale=1]
%uncomment if require: \path (0,163); %set diagram left start at 0, and has height of 163

%Straight Lines [id:da501871060920984] 
\draw [color={rgb, 255:red, 74; green, 144; blue, 226 }  ,draw opacity=1 ]   (98,94.92) -- (97.6,134.67) ;
%Straight Lines [id:da8259162856057432] 
\draw [color={rgb, 255:red, 74; green, 144; blue, 226 }  ,draw opacity=1 ][fill={rgb, 255:red, 74; green, 144; blue, 226 }  ,fill opacity=1 ]   (18.56,15.06) -- (31.8,28.37) -- (45.04,41.68) -- (58.28,54.99) -- (71.52,68.3) -- (84.76,81.61) -- (98,94.92) ;
%Straight Lines [id:da1645661736945132] 
\draw [color={rgb, 255:red, 74; green, 144; blue, 226 }  ,draw opacity=1 ]   (167.56,14.72) -- (98,94.92) ;
%Straight Lines [id:da03058805914013152] 
\draw [color={rgb, 255:red, 74; green, 144; blue, 226 }  ,draw opacity=1 ]   (88.22,45.39) -- (68.22,65.06) ;
%Straight Lines [id:da8401862944408585] 
\draw [color={rgb, 255:red, 74; green, 144; blue, 226 }  ,draw opacity=1 ]   (88.22,14.72) -- (88.22,45.39) ;
%Straight Lines [id:da22319743847914375] 
\draw [color={rgb, 255:red, 74; green, 144; blue, 226 }  ,draw opacity=1 ]   (302.74,94.76) -- (303.14,134.42) ;
%Straight Lines [id:da5128871303098825] 
\draw [color={rgb, 255:red, 74; green, 144; blue, 226 }  ,draw opacity=1 ]   (382.33,15.1) -- (302.74,94.76) ;
%Straight Lines [id:da5795657849845721] 
\draw [color={rgb, 255:red, 74; green, 144; blue, 226 }  ,draw opacity=1 ]   (233.06,14.77) -- (302.74,94.76) ;
%Straight Lines [id:da3875529114818561] 
\draw [color={rgb, 255:red, 74; green, 144; blue, 226 }  ,draw opacity=1 ]   (312.54,45.36) -- (332.57,64.98) ;
%Straight Lines [id:da22771677986583327] 
\draw [color={rgb, 255:red, 74; green, 144; blue, 226 }  ,draw opacity=1 ]   (312.54,14.77) -- (312.54,45.36) ;
%Left Right Arrow [id:dp44901680864746485] 
\draw   (180.8,125.1) -- (190.59,120.29) -- (190.59,122.69) -- (210.16,122.69) -- (210.16,120.29) -- (219.95,125.1) -- (210.16,129.92) -- (210.16,127.51) -- (190.59,127.51) -- (190.59,129.92) -- cycle ;
%Shape: Circle [id:dp8519736673437364] 
\draw  [color={rgb, 255:red, 208; green, 2; blue, 27 }  ,draw opacity=1 ][fill={rgb, 255:red, 208; green, 2; blue, 27 }  ,fill opacity=1 ] (251.46,84.86) .. controls (251.46,83.88) and (252.26,83.08) .. (253.24,83.08) .. controls (254.22,83.08) and (255.02,83.88) .. (255.02,84.86) .. controls (255.02,85.84) and (254.22,86.64) .. (253.24,86.64) .. controls (252.26,86.64) and (251.46,85.84) .. (251.46,84.86) -- cycle ;

% Text Node
\draw (7.89,10.5) node [anchor=north west][inner sep=0.75pt]  [font=\scriptsize,color={rgb, 255:red, 74; green, 144; blue, 226 }  ,opacity=1 ]  {$g$};
% Text Node
\draw (76.81,10.5) node [anchor=north west][inner sep=0.75pt]  [font=\scriptsize,color={rgb, 255:red, 74; green, 144; blue, 226 }  ,opacity=1 ]  {$h$};
% Text Node
\draw (155.76,10.5) node [anchor=north west][inner sep=0.75pt]  [font=\scriptsize,color={rgb, 255:red, 74; green, 144; blue, 226 }  ,opacity=1 ]  {$k$};
% Text Node
\draw (88.24,139.07) node [anchor=north west][inner sep=0.75pt]  [font=\scriptsize,color={rgb, 255:red, 74; green, 144; blue, 226 }  ,opacity=1 ]  {$ghk$};
% Text Node
\draw (222.02,10.5) node [anchor=north west][inner sep=0.75pt]  [font=\scriptsize,color={rgb, 255:red, 74; green, 144; blue, 226 }  ,opacity=1 ]  {$g$};
% Text Node
\draw (302.56,10.5) node [anchor=north west][inner sep=0.75pt]  [font=\scriptsize,color={rgb, 255:red, 74; green, 144; blue, 226 }  ,opacity=1 ]  {$h$};
% Text Node
\draw (371.51,10.5) node [anchor=north west][inner sep=0.75pt]  [font=\scriptsize,color={rgb, 255:red, 74; green, 144; blue, 226 }  ,opacity=1 ]  {$k$};
% Text Node
\draw (294.17,138.82) node [anchor=north west][inner sep=0.75pt]  [font=\scriptsize,color={rgb, 255:red, 74; green, 144; blue, 226 }  ,opacity=1 ]  {$ghk$};
% Text Node
\draw (231.68,89.26) node [anchor=north west][inner sep=0.75pt]  [font=\scriptsize,color={rgb, 255:red, 208; green, 2; blue, 27 }  ,opacity=1 ]  {$\beta ( g,h,k)$};
% Text Node
\draw (194.88,132.79) node [anchor=north west][inner sep=0.75pt]   [align=left] {F};
% Text Node
\draw (61.72,77.01) node [anchor=north west][inner sep=0.75pt]  [font=\scriptsize,color={rgb, 255:red, 74; green, 144; blue, 226 }  ,opacity=1 ]  {$gh$};
% Text Node
\draw (326.07,77.01) node [anchor=north west][inner sep=0.75pt]  [font=\scriptsize,color={rgb, 255:red, 74; green, 144; blue, 226 }  ,opacity=1 ]  {$hk$};

\end{tikzpicture}

\end{center}

This represents diagrammatically a 2-group symmetry $\Gamma$ whose gauging one can consider. Physically, gauging the 2-group $\Gamma$ encodes a gauging of both the ordinary group $G$ as well
as $BK$, so that the path integral contains sums over both $G$ bundles as well as $K$-gerbes (each of which is twisted by the other, in a fashion encoded by $\alpha$, $\beta_{\Gamma}$, as represented in the illustrations above).

In this paper, we will utilize presentations of 2-groups as collections $(\Gamma_1, \Gamma_0, \delta, \overline{\alpha})$,
where $\Gamma_{0,1}$ are (not necessarily abelian) groups, 
and $\delta: \Gamma_1 \rightarrow \Gamma_0$ and
$\overline{\alpha}: \Gamma_0 \rightarrow {\rm Aut}(\Gamma_1)$ are homomorphisms satisfying identities
listed in appendix~\ref{app:background}.  This structure is known as a crossed module.  In terms of such
presentations, $G = {\rm coker}\, \delta$, $K = \ker \delta$,
$\beta_{\Gamma} \in H^3(G,K)$ is determined by the class of the
exact sequence
\begin{equation}
    1 \: \longrightarrow \: K \: \longrightarrow \: \Gamma_1
    \: \stackrel{\delta}{\longrightarrow} \: \Gamma_0 \: \longrightarrow \: G \: \longrightarrow \: 1,
\end{equation}
where $K$ is an abelian group, and $\alpha$ is determined by $\overline{\alpha}$ (See appendix~\ref{app:background} for further details of crossed modules).
The notation 
\begin{equation}
    {}^g k \: = \: \overline{\alpha}(g)(k),
\end{equation}
for $k \in \Gamma_1$ and $g \in \Gamma_0$,
is sometimes also used to indicate the $\overline{\alpha}$ action of $\Gamma_0$ (as well as the $\alpha$ action of $G$ on $K$).  Also, a crossed module presentation of $\Gamma$ is often denoted
\begin{equation}
    \Gamma \: = \: \left( \Gamma_1 \stackrel{\delta}{\longrightarrow} \Gamma_0 \right),
\end{equation}
with $\overline{\alpha}$ given separately.

A given 2-group $\Gamma$ might have many different crossed module presentations $(\Gamma_1, \Gamma_0,
\delta, \overline{\alpha})$; however, $G$ and $K$
are independent of presentation,
as is the
homomorphism $\alpha: G \rightarrow {\rm Aut}(K)$ and
the class in $\beta_{\Gamma} \in H^3(G,K)$.

The action of the 2-group $\Gamma$ on a space $X$ factors through $G$, meaning, in terms of a presentation
$(\Gamma_1, \Gamma_0, \delta, \overline{\alpha})$,
it is encoded
in a $\Gamma_0$ action on $X$ such that the image of $\delta$ acts trivially on $X$.
(See appendix~\ref{ssec:xmod} for details.)  The $K$-gerbe acts trivially on the space. One can consider more general scenarios in which $X$ is not just a smooth manifold but a differentiable stack such as the moduli stack of a Lie groupoid. In such a case, the gerbe is allowed to have a nontrivial action on the target.

The fact that one gauges a $1$-form symmetry $BK$ in \textit{three} dimensions that acts trivially on the space is crucial to observe decomposition. In \cite{Sharpe:2019ddn}, in \textit{two} dimensions, these higher morphisms translated to $(d-1)$-form symmetries on the worldsheet, which is ultimately why gauging the whole 2-group only returns the theory of an orbifold by the cokernel. In contrast, the higher morphisms here represent $(d-2)$-form symmetries on the worldvolume, so we expect to see a less trivial decomposition in this case.

Graphically, before gauging, each of the lines associated with elements of the
$BK$ can end on a (dimension-zero) defect operator, which after gauging can float free.   Defect operators originating as endpoints of trivially-acting lines have weight zero, and can be combined to form projectors, following
Wedderburn's theorem.

We remind the reader that the higher morphisms will necessarily act trivially on the target space.  As explored in \cite{Sharpe:2019ddn}, in 2d, these higher morphisms translated to $(d-1)$-form symmetries on the worldsheet, which is ultimately why gauging the whole 2-group only returns the theory of an orbifold by the cokernel.  In contrast, the higher morphisms here will represent $(d-2)$-form symmetries on the worldvolume, so we expect to see a less trivial decomposition in this case.

In the remainder of this section, we will discuss basic properties of three-dimensional orbifolds
by 2-groups, generalizing \cite{Pantev:2022kpl}.

\subsection{Torus partition function}\label{ssec:3tor}

As in previous papers, we work with partition functions. In this section, we describe the computation of the partition function of a 2-group orbifold on a 3-torus $T^3$. 
Our analysis will generalize the results of \cite{Pantev:2022kpl}.

First, we consider the partition function of the 2-group orbifold $[X/\Gamma]$,
for $\Gamma$ presented as the crossed module $(\Gamma_1\to\Gamma_0)$, where $\Gamma_0$ and $\Gamma_1$ are both finite.
Intuitively, since $[X/\Gamma]$ describes a gauging of both the 0-form symmetry $\Gamma_0$
as well as the one-form symmetry $\Gamma_1$,
the path integral may be regarded as summing over both $\Gamma_0$ bundles on $T^3$ as well as certain gerbes on the $\Gamma_0$ quotient.
This is, of course, just an intuitive picture, since in reality $\Gamma_1$ and $\Gamma_0$ twist each other. In principle, the only structure that has meaning is that of a principal $\Gamma$-bundle.  Taking into account the overall factors reflecting gauge volumes
described in
\cite[section 4.1.3]{Pantev:2022kpl}, the partition function on $T^3$ should have the form
\begin{equation}
Z_{T^3}\left( [X/\Gamma] \right) \: = \:
\frac{ | H^0(T^3,\Gamma_1) | }{ | H^1(T^3, \Gamma_1 )| } \frac{1}{| H^0(T^3, \Gamma_0) | }
 \sideset{}{'} \sum_{z_1, z_2, z_3 \in \Gamma_1} \sideset{}{'} \sum_{g,h,k \in \Gamma_0}
Z(g,h,k; z_1,z_2,z_3),
\end{equation}
where the prime on the sums indicate that $g, h, k, z_1, z_2, z_3$ need to satisfy some conditions which we will determine shortly.
The sum over $z_{1,2,3}$ is regarded as a sum over equivalence classes of banded $\Gamma_1$-gerbes, classified by
\begin{equation}
    H^2(T^3, \Gamma_1) \: = \: \left( \Gamma_1 \right)^3
\end{equation}
(for $\Gamma_1$ finite),
and the sum over $g, h, k$ as a sum over equivalence classes of ordinary $\Gamma_0$-bundles, classified by
\begin{equation}
    H^1(T^3, \Gamma_0) \: = \: \left( \Gamma_0 \right)^3
\end{equation}
(for $\Gamma_0$ finite). It is of course the tuple $(g,h,k;z_1,z_2,z_3)$ that determines a principal $\Gamma$-bundle. Although discrete torsion can be added, for the moment we are focused on orbifold partition functions without discrete torsion.

Now, let us turn to the conditions that the group elements above must satisfy.
Since all the groups acting on the target space $X$ are finite, we will only deal with flat $\Gamma$-principal bundles. As described in \cite{porter1,porter2}, these bundles can be specified by triangulating the worldvolume and endowing such simplices with $\Gamma$-colorings described by a simplicial map from the triangulation $S(T^3)$ of $T^3$ to $B\Gamma$. In the usual case for $\Gamma=G$ a group, it is well-known that principal $G$-bundles on $T^3$ can be described by triples $(g,h,k)\in G^3$ of pairwise commuting elements. This is obtained from the simplicial map as follows. For 0-simplices, or vertices, $BG$ only has one, so trivial information is assigned to vertices in $S(T^3)$. The 1-simplices of $BG$ are group elements, so we assign group elements to each edge in $S(T^3)$. Clearly, the assignments for each edge in the triangulation can be deduced from the assignments to the generators of $\pi_1(T^3)=\Z^3$. Moreover, since the only information that 2-simplices in $BG$ contain is about composition of group elements, this forces the group elements assigned to the edges of $S(T^3)$ to pairwise commute. 

On the other hand, if $\xmod$ is a 2-group, a strict model for a flat principal $\Gamma$ bundle can be described by assigning  group elements of $\Gamma_0$ to 1-simplices, and group elements of $\Gamma_1\rtimes\Gamma_0$ to 2-simplices, and since $B\Gamma$ does not have nontrivial 3-morphisms, then this induces a consistency condition on the $\Gamma_1$ assignments deriveable from the simplicial identities. This says that the $\Gamma$-bundle is determined by a triple of group elements $(g,h,k)\in (\Gamma_0)^3$ such that, instead of pairwise commuting, only commute up to the action of $(z_1,z_2,z_3)\in(\Gamma_1)^3$, which in turn satisfy a gluing condition. The conditions are listed in Eqs.~(\ref{eq:3dconditions})-(\ref{eq:3dconditions4}) below.

Altogether, the partition function for $T^3$ on an orbifold by a 2-group $\Gamma=(\Gamma_1\to\Gamma_0)$ (without discrete torsion) is
\begin{equation*}
    Z_{T^3}([X/\Gamma]) = \frac{1}{|\Gamma_1|^2 |\Gamma_0|} \sideset{}{'}\sum_{g,h,k\in \Gamma_0;\ z_1,z_2,z_3\in \Gamma_1} Z(g,h,k;z_1,z_2,z_3)
\end{equation*}
such that $g,h,k\in \Gamma_0$ and $z_1,z_2,z_3\in \Gamma_1$ satisfy the constraints
\begin{eqnarray}\label{eq:3dconditions}
    gh=\delta(z_1)hg, 
    \\ 
    gk=\delta(z_2)kg,\label{eq:3dconditions2}
    \\
    hk=\delta(z_3)kh,\label{eq:3dconditions3}
    \\ 
    z_1 \tensor[^h]{z}{_2}z_3 = \tensor[^g]{z}{_3}z_2\tensor[^k]{z}{_1}. \label{eq:3dconditions4}
\end{eqnarray}
Morally, the sum over elements of $\Gamma_0$ is the path integral's sum over $\Gamma_0$ bundles, and 
%at least when $\Gamma_1$ is abelian,
the sum over elements of $\Gamma_1$ is the path integral's sum over certain gerbes over the quotient by $\Gamma_0$.
%(There are three of each because on a $T^3$, $H^1(T^3,K) = H^2(T^3,K)$ for any abelian group $K$.) 
The overall factors arise for the same reasons as in \cite[section 4.1.3]{Pantev:2022kpl}.

For simplicity, we will denote this partition function as
\begin{equation}\label{eq:partftn}
    Z_{T^3}([X/\Gamma]) = \frac{1}{|\Gamma_1|^2|\Gamma_0|}\sum_{(g,h,k;z_1,z_2,z_3)\in\Gamma} Z(g,h,k;z_1,z_2,z_3)
\end{equation}
where the sum is understood as summing over group elements $g,h,k\in\Gamma_0$ and $z_1,z_2,z_3\in\Gamma_1$ satisfying the four constraints above.

So far, we have presented the partition function in terms
of the data defining a presentation of the 2-group.  Such presentations are not unique.  Nevertheless, 
the partition function
is well-defined, because it can be recast in terms of a computation that involves only $G$, $K$, $\alpha:G\to\text{Aut}(K)$, and $\beta_{\Gamma}\in H^3(G,K)$. This is done in close analogy to what is presented in \cite{Pantev:2022kpl}. 
Let ${\cal K}_{\alpha}$ be the bundle of abelian groups on $BG$ corresponding to $\alpha$, that is the bundle of abelian groups associated with the principal $G$ bundle ${\rm pt} \rightarrow BG$ by $\alpha$.
Let $M$ be the worldvolume, and $x:M\to BG$ any map. By the fibration
\begin{equation}
    \cdots\longrightarrow B^2{\cal K}_{\alpha}\longrightarrow B\Gamma\longrightarrow BG \stackrel{\beta_{\Gamma}}{\longrightarrow} B^3 {\cal K}_{\alpha}\longrightarrow\cdots 
\end{equation}
all $G$-twisted sectors (bundles) $Z(g,h,k)$, each represented by a map $x$, satisfying 
\begin{displaymath}
    x^*\beta_{\Gamma}(g,h,k)=1,
\end{displaymath}
lift to a $\Gamma$-twisted sector $Z(g',h',k';z_1,z_2,z_3)$, each which is a map $M\to B\Gamma$. Any lift of $g,h,k$ is up to an element in $\text{Im}(\delta)$, so that there is a $|\text{Im}(\delta)|^3$-fold ambiguity in these lifts. Moreover, the elements $z_1,z_2,z_3$ could each be multiplied by elements $k_1,k_2,k_3$ in $K=\kerd$ and still satisfy the first three constraints (\ref{eq:3dconditions})-(\ref{eq:3dconditions3}) that define a $\Gamma$-bundle. However, $Z(g',h',k';k_1z_1,k_2z_2,k_3z_3)$ is another $\Gamma$-twisted sector only if $k_1\tensor[^{h'}]{k}{_2}k_3=\tensor[^{g'}]{k}{_3}k_2\tensor[^{k'}]{k}{_1}$, since $k_1,k_2,k_3\in K\subset Z(\Gamma_1)$. Note that if these $k$'s satisfy such an equation for a given lift $g',h',k'$, then they satisfy the equation for any other choice of lift. This is because the lifts only differ by elements in $\text{Im}(\delta)$, which by the crossed module identities act trivially on $Z(\Gamma_1)$. This exhausts all possibilities of $\Gamma$-twisted sectors. For the particular case of $M=T^3$, the condition $x^*\omega=1$ is precisely the statement that the associated discrete torsion-like phase valued in $K$ is trivial, that is
\begin{equation}
    \epsilon_{\beta_{\Gamma}}(g,h,k)=\frac{\beta_{\Gamma}(g,h,k)}{\beta_{\Gamma}(g,k,h)}\frac{\beta_{\Gamma}(k,g,h)}{\beta_{\Gamma}(k,h,g)}\frac{\beta_{\Gamma}(h,k,g)}{\beta_{\Gamma}(h,g,k)}=1\in K.
\end{equation}

Therefore
\begin{eqnarray}
    Z_{T^3}([X/\Gamma]) & = &
    \frac{1}{|\Gamma_1|^2|\Gamma_0|}\sum_{(g,h,k;z_1,z_2,z_3)\in\Gamma} Z(g,h,k;z_1,z_2,z_3), 
    \\
    & = & \frac{|\text{Im}(\delta)|^3}{|\Gamma_1|^2 |\Gamma_0|} \sideset{}{'}\sum_{k_1,k_2,k_3\in K} \sideset{}{'}\sum_{g,h,k\in G} Z(g,h,k;k_1,k_2,k_3), 
    \\
    & = & \frac{1}{|K|^2 |G|}\sideset{}{'}\sum_{k_1,k_2,k_3\in K}\sideset{}{'}\sum_{g,h,k\in G} Z(g,h,k;k_1,k_2,k_3),
    \label{eq:Z:invt}
\end{eqnarray}
where the sum is over tuples $(g,h,k;k_1,k_2,k_3)$ such that $g,h,k\in G$ pairwise commute and satisfy $\epsilon_{\beta_{\Gamma}}(g,h,k)=1\in K$, and $k_1,k_2,k_3\in K$ satisfy $k_1\tensor[^h]{k}{_2}k_3=\tensor[^g]{k}{_3}k_2\tensor[^k]{k}{_1}$. In terms of defect diagrams, the latter amounts to the condition:
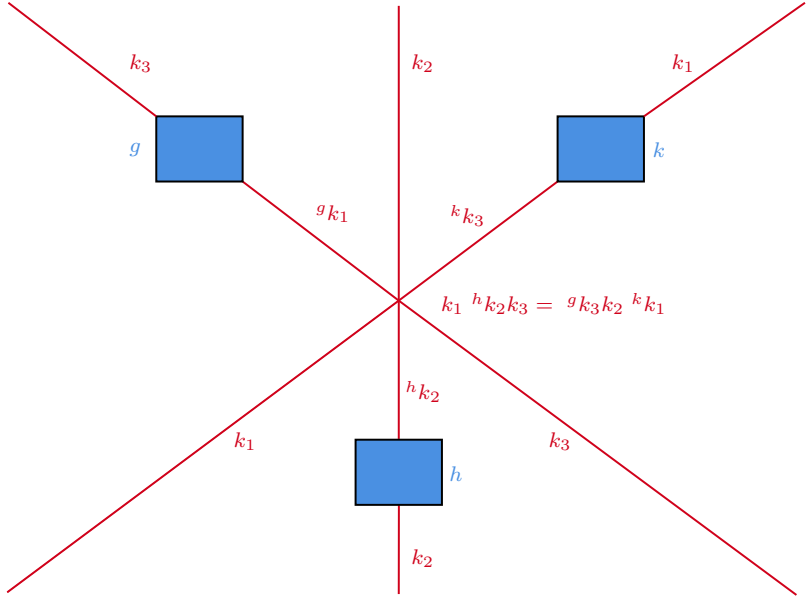
\begin{center}

\tikzset{every picture/.style={line width=0.75pt}} %set default line width to 0.75pt        

\begin{tikzpicture}[x=0.75pt,y=0.75pt,yscale=-1,xscale=1]

%Straight Lines [id:da05816526423581547] 
\draw [color={rgb, 255:red, 208; green, 2; blue, 27 }  ,draw opacity=1 ][fill={rgb, 255:red, 208; green, 2; blue, 27 }  ,fill opacity=1 ]   (349.67,151) -- (152.22,298.22) ;
%Straight Lines [id:da24015322057263444] 
\draw [color={rgb, 255:red, 208; green, 2; blue, 27 }  ,draw opacity=1 ]   (349.67,151) -- (550.22,299.56) ;
%Straight Lines [id:da6790604463348748] 
\draw [color={rgb, 255:red, 208; green, 2; blue, 27 }  ,draw opacity=1 ]   (270.92,90.89) -- (349.67,151) ;
%Straight Lines [id:da12662996786715852] 
\draw [color={rgb, 255:red, 208; green, 2; blue, 27 }  ,draw opacity=1 ]   (429.77,90.89) -- (349.67,151) ;
%Straight Lines [id:da4672266137650256] 
\draw [color={rgb, 255:red, 208; green, 2; blue, 27 }  ,draw opacity=1 ]   (349.67,2.22) -- (349.67,151) ;
%Straight Lines [id:da5172570885186927] 
\draw [color={rgb, 255:red, 208; green, 2; blue, 27 }  ,draw opacity=1 ]   (349.73,254.27) -- (349.67,299.11) ;
%Straight Lines [id:da17294011633308104] 
\draw [color={rgb, 255:red, 208; green, 2; blue, 27 }  ,draw opacity=1 ]   (349.67,151) -- (349.67,220.8) ;
%Shape: Rectangle [id:dp5723264288536014] 
\draw  [fill={rgb, 255:red, 74; green, 144; blue, 226 }  ,fill opacity=1 ] (227.37,58) -- (270.92,58) -- (270.92,90.89) -- (227.37,90.89) -- cycle ;
%Shape: Rectangle [id:dp7127590405044018] 
\draw  [fill={rgb, 255:red, 74; green, 144; blue, 226 }  ,fill opacity=1 ] (327.89,221.2) -- (371.44,221.2) -- (371.44,254.09) -- (327.89,254.09) -- cycle ;
%Straight Lines [id:da6898032841900299] 
\draw [color={rgb, 255:red, 208; green, 2; blue, 27 }  ,draw opacity=1 ]   (152.8,0.67) -- (227.37,58) ;
%Shape: Rectangle [id:dp79126445062492] 
\draw  [fill={rgb, 255:red, 74; green, 144; blue, 226 }  ,fill opacity=1 ] (429.77,58) -- (473.32,58) -- (473.32,90.89) -- (429.77,90.89) -- cycle ;
%Straight Lines [id:da8297195387305596] 
\draw [color={rgb, 255:red, 208; green, 2; blue, 27 }  ,draw opacity=1 ]   (554.53,0.67) -- (473.32,58) ;

% Text Node
\draw (264.92,216.1) node [anchor=north west][inner sep=0.75pt]  [font=\scriptsize,color={rgb, 255:red, 208; green, 2; blue, 27 }  ,opacity=1 ]  {$k_{1}$};
% Text Node
\draw (423.77,216.1) node [anchor=north west][inner sep=0.75pt]  [font=\scriptsize,color={rgb, 255:red, 208; green, 2; blue, 27 }  ,opacity=1 ]  {$k_{3}$};
% Text Node
\draw (354.57,274.9) node [anchor=north west][inner sep=0.75pt]  [font=\scriptsize,color={rgb, 255:red, 208; green, 2; blue, 27 }  ,opacity=1 ]  {$k_{2}$};
% Text Node
\draw (351.57,189.8) node [anchor=north west][inner sep=0.75pt]  [font=\scriptsize,color={rgb, 255:red, 208; green, 2; blue, 27 }  ,opacity=1 ]  {$^{h} k_{2}$};
% Text Node
\draw (369.57,145.4) node [anchor=north west][inner sep=0.75pt]  [font=\scriptsize,color={rgb, 255:red, 208; green, 2; blue, 27 }  ,opacity=1 ]  {$k_{1} \ ^{h} k_{2} k_{3} =\ ^{g} k_{3} k_{2} \ ^{k} k_{1} \ $};
% Text Node
\draw (373.77,233.04) node [anchor=north west][inner sep=0.75pt]  [font=\scriptsize,color={rgb, 255:red, 74; green, 144; blue, 226 }  ,opacity=1 ]  {$h$};
% Text Node
\draw (212.37,69.84) node [anchor=north west][inner sep=0.75pt]  [font=\scriptsize,color={rgb, 255:red, 74; green, 144; blue, 226 }  ,opacity=1 ]  {$g$};
% Text Node
\draw (476.32,69.84) node [anchor=north west][inner sep=0.75pt]  [font=\scriptsize,color={rgb, 255:red, 74; green, 144; blue, 226 }  ,opacity=1 ]  {$k$};
% Text Node
\draw (212.37,25.4) node [anchor=north west][inner sep=0.75pt]  [font=\scriptsize,color={rgb, 255:red, 208; green, 2; blue, 27 }  ,opacity=1 ]  {$k_{3}$};
% Text Node
\draw (306.33,101.15) node [anchor=north west][inner sep=0.75pt]  [font=\scriptsize,color={rgb, 255:red, 208; green, 2; blue, 27 }  ,opacity=1 ]  {$^{g} k_{1}$};
% Text Node
\draw (485.97,25.4) node [anchor=north west][inner sep=0.75pt]  [font=\scriptsize,color={rgb, 255:red, 208; green, 2; blue, 27 }  ,opacity=1 ]  {$k_{1}$};
% Text Node
\draw (374.12,101.15) node [anchor=north west][inner sep=0.75pt]  [font=\scriptsize,color={rgb, 255:red, 208; green, 2; blue, 27 }  ,opacity=1 ]  {$^{k} k_{3}$};
% Text Node
\draw (354.57,25.4) node [anchor=north west][inner sep=0.75pt]  [font=\scriptsize,color={rgb, 255:red, 208; green, 2; blue, 27 }  ,opacity=1 ]  {$k_{2}$};

\end{tikzpicture}

\end{center}

This shows that the partition function can be computed in terms of the topological data that defines $\Gamma$, so that it does not depend on the particular presentation $\xmod$. In practice, however, it will often be more straightforward to work with Equation~(\ref{eq:partftn}), since the evaluation of $\beta_{\Gamma}(g,h,k)$ is often cumbersome. Nevertheless, this expression is useful whenever an explicit presentation of $\Gamma$ in terms of a crossed module is not readily available.

So far, we have discussed partition functions without (analogues of)
discrete torsion.  Now, there does exist an analogue of discrete
torsion for three-dimensional two-group orbifolds, as we shall
discuss in section~\ref{sec:3ddt}.  We discuss there the phases that arise in partition functions, and later in section~\ref{sect:decomp-dt}
will outline the ramifications for decomposition; however,
our focus in this paper is primarily on orbifolds without 2-group discrete torsion.

\subsection{More general 3d worldvolumes}

Even though we have focused on obtaining the partition function for when the worldvolume is a 3-torus, more general 3d manifolds can be considered. As previously described, the process described above relies on two facts. First, that the worldvolume admits a triangulation. Indeed, it is known that any smooth manifold admits a triangulation (see e.g. \cite{manolescu2014triangulations}). The second fact is that principal bundles with flat connections can be described in terms of holonomies (see e.g.~\cite{taubes}), or simplicial maps \cite{porter1,porter2}. These two results allow us to describe the $\Gamma$-bundle, which is necessarily flat, in terms of a simplicial map from the triangulation of the worldsheet to the simplicial presentation of the classifying space $B\Gamma$, from where we can derive gluing conditions analogous to Eq.~(\ref{eq:3dconditions})-(\ref{eq:3dconditions4}). 

This is by now a standard process that arose in the context of Dijkgraaf-Witten theory \cite{Dijkgraaf:1989pz} and its generalizations (such as the Yetter model \cite{yetter}), made precise by T. Porter in \cite{porter1,porter2}. Such theories are described by $\sigma$-models with homotopy $n$-types as target spaces, which correspond to the classifying spaces of $n$-groups $\Gamma$. It was observed that those models, including analogues of discrete torsion, can be described as $\Gamma$-colorings of triangulations of the worldvolume. As such, those models may be regarded as specializations of the situation described presently, where $X=*$ and so $[*/\Gamma]=B\Gamma$ (i.e.~all of $\Gamma$ acts trivially), so that the decomposition results presented here in particular apply to those models.  

Morally, to each 1-simplex in the triangulation of the worldvolume, we assign an element of $\Gamma_0$, and to each 2-simplex we assign an element of $\Gamma_1\rtimes \Gamma_0$. The simplicial identities of the map enforce conditions on these assignments. For example, for a simplex \textbf{012} with $g,h,k\in\Gamma_0$ assigned to \textbf{01}, \textbf{12}, and \textbf{02}, respectively, and $(z,l)\in\Gamma_1\rtimes \Gamma_0$ assigned to \textbf{012}, we require that $l=hg$ and that $k=\delta(z)hg$. Since the 2-group $\Gamma$ has $\pi_3(B\Gamma)=0$, then trivial information is assigned to every 3-simplex. This will imply gluing conditions on the elements in $\Gamma_1\rtimes \Gamma_0$ assigned to the 2-simplices, such as the fourth equation in Eq.~(\ref{eq:3dconditions}), which morally says that the surfaces enclose a closed volume with trivial information assigned to it. Of course, if $\Gamma$ was a 3-group, then we would also assign a group element (or 3-morphism) to 3-simplices, which, for example, would weaken the equations on 2-simplices. For example, in the case of the 3-torus, we would end up with an \textit{isomorphism}
\begin{equation}
    z_1 \tensor[^h]{z}{_2}z_3 \Longrightarrow \tensor[^g]{z}{_3}z_2\tensor[^k]{z}{_1}
\end{equation}
realized by a 3-morphism in $\Gamma$.

As for discrete torsion, we will have that the map $B\Gamma\to B^3U(1)$ will result in assigning a 3-cocycle classified by $H^3(\Gamma,U(1))$ to each 3-simplex in the worldvolume triangulation. The resulting discrete torsion phase is then obtained by taking the alternating product of these cocycles. This is briefly illustrated in Section \ref{sssec:phases} for the 3-torus. Therefore, the process, albeit involved, becomes purely combinatorial, and more importantly, generalizes both to higher-dimensional worldvolumes, and higher groups.

\subsection{A comment on trivially-acting groups}\label{ssec:trivially-acting}

 Whenever there is a subgroup of the orbifold group that acts trivially, the twisted sectors that are related by elements of the such subgroup are the same. This was used, for example, in \cite{Pantev:2005rh} in the context of a 2d $[X/G]$ orbifold with $A<G$ a trivially-acting subgroup. Because $A$ acts trivially, then if $Z(g,h)$ and $Z(ag,a'h)$ are both twisted sectors in the $[X/G]$, then $Z(g,h)=Z(ag,a'h)$, and so we can just label the twisted sectors by the equivalence classes in $G/A$ as $Z([g],[h])$.

In Section~\ref{ssec:2grpact}, we observe that higher morphisms (i.e.~elements of $\Gamma_1$ in $\xmod$) necessarily act trivially on the target space $X$. Therefore, much of the labeling of twisted sectors $Z(g,h,k;z_1,z_2,z_3)$ in the $[X/\Gamma]$ theory is redundant. Ultimately, the physical information will be obtained after quotienting everything down to $\cokerd$, which is the only part of $\Gamma$ that acts non-trivially on $X$. Thus, physically, the projection
\begin{center}
    \begin{tikzcd}
\Gamma_1 \arrow[r, "\delta"] \arrow[d, "0"'] & \Gamma_0 \arrow[d, "\pi"] \\
1 \arrow[r, "0"']                            & \text{coker}(\delta)     
\end{tikzcd}
\end{center}
will only remove the redundant labelling (in other words, will sum the twisted sectors that are the same) but leave the physical information intact. 

Note that the commutative diagram presented above is actually a homomorphism of crossed modules, which we call $\Pi:\Gamma\to \cokerd$, and thus in particular induces a map $\Pi^*: H^3(\cokerd,U(1))\to H^3(\Gamma,U(1))$. It is from this observation that we derive part of our story.

\subsection{2-group analogues of discrete torsion}\label{sec:3ddt}

In this section, we will briefly outline 2-group analogues of discrete torsion, classified by
$H^3(\Gamma,U(1)) = H^3_{\rm sing}(B\Gamma,U(1))$.
They will not play a substantial role in this paper, so our discussion will be brief.

\subsubsection{Phases}\label{sssec:phases}

For ordinary group orbifolds in three dimensions, there is an analogue of discrete torsion, counted by $H^3(G,U(1))$, see e.g.~\cite{Dijkgraaf:1989pz,Sharpe:2000qt,Pantev:2022kpl}.
For example, on a $T^3$, the discrete torsion phase associated to a $(g,h,k)$ twisted sector is
\begin{equation}
    \frac{\omega(g,h,k)}{\omega(g,k,h)}\cdot \frac{\omega(k,g,h)}{\omega(k,h,g)}\cdot \frac{\omega(h,k,g)}{\omega(h,g,k)}
\end{equation}
where $[\omega]\in H^3(G,U(1))$. (Although the expression is written in terms of a cocycle, it is easily checked to be invariant under coboundaries -- hence well defined on group cohomology -- and in addition is $SL(3,{\mathbb Z})$ invariant.)
Just as in the 2d case, we can see that these phases exactly come from the simplices that compose the torus. Indeed, the triple $(g,h,k)$ can be regarded as describing a 3-simplex such that $\omega(g,h,k)$ is the $U(1)$-valued cocycle assigned to it in degree-3 group cohomology.

In close analogy, 2-group discrete torsion phases are given by the product of 3-cocycles in $H^3(\Gamma,U(1))$, such that the order of multiplication is given by the ordering of the 3-simplices that compose the 3-torus, essentially obtained by the description in \cite{porter1,porter2}. For example, the usual phase $\omega(g,k,h)$ becomes in this case $\omega(g,k,h; z_2,z_1,z_3^{-1})$, which is the $U(1)$ phase assigned to the 3-simplex:
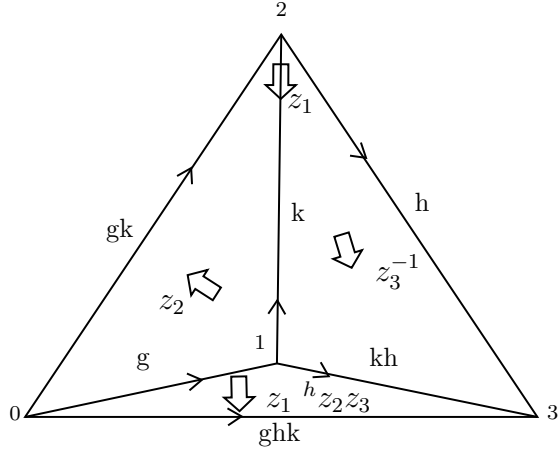
\begin{center}
    \begin{tikzcd}

\tikzset{every picture/.style={line width=0.75pt}} %set default line width to 0.75pt        

\begin{tikzpicture}[x=0.75pt,y=0.75pt,yscale=-1,xscale=1]

%Shape: Triangle [id:dp11310657279891112] 
\draw   (322.17,43) -- (451.33,236) -- (193,236) -- cycle ;
%Straight Lines [id:da7068287779192839] 
\draw    (322.17,43) -- (320,209) ;
%Straight Lines [id:da6566800578490517] 
\draw    (320,209) -- (193,236) ;
%Straight Lines [id:da8515469563761646] 
\draw    (451.33,236) -- (320,209) ;
\draw   (274.6,214.1) -- (282.09,217.32) -- (275.88,222.6) ;
\draw   (295.04,231.67) -- (302.13,235.7) -- (295.37,240.26) ;
\draw   (315.64,183.94) -- (320.19,177.17) -- (324.24,184.25) ;
\draw   (269.63,114.4) -- (276.73,110.39) -- (277.17,118.54) ;
\draw   (362.35,97.51) -- (364.93,105.25) -- (356.85,104.12) ;
\draw   (342.06,208.4) -- (346.12,215.48) -- (337.97,215.97) ;
%Up Arrow [id:dp25742103431806007] 
\draw   (308.58,225.93) -- (301.13,233.1) -- (293.37,226.27) -- (297.17,226.19) -- (296.94,215.68) -- (304.54,215.52) -- (304.78,226.02) -- cycle ;
%Up Arrow [id:dp6381130536306536] 
\draw   (276.75,174.6) -- (274.98,164.41) -- (284.98,161.81) -- (282.92,165.01) -- (291.75,170.69) -- (287.64,177.09) -- (278.81,171.4) -- cycle ;
%Up Arrow [id:dp0018474591442700916] 
\draw   (362.92,151.81) -- (357.69,160.72) -- (348.37,156.24) -- (352.01,155.13) -- (348.95,145.08) -- (356.22,142.87) -- (359.28,152.92) -- cycle ;
%Up Arrow [id:dp7522787672691225] 
\draw   (329.53,68.53) -- (321.9,75.5) -- (314.32,68.47) -- (318.13,68.49) -- (318.17,57.98) -- (325.77,58.01) -- (325.73,68.52) -- cycle ;

% Text Node
\draw (247.6,202.53) node [anchor=north west][inner sep=0.75pt]   [align=left] {{\footnotesize g}};
% Text Node
\draw (388,121.33) node [anchor=north west][inner sep=0.75pt]   [align=left] {{\footnotesize h}};
% Text Node
\draw (325.6,126.53) node [anchor=north west][inner sep=0.75pt]   [align=left] {{\footnotesize k}};
% Text Node
\draw (232.4,134.53) node [anchor=north west][inner sep=0.75pt]   [align=left] {{\footnotesize gk}};
% Text Node
\draw (309.2,237.33) node [anchor=north west][inner sep=0.75pt]   [align=left] {{\footnotesize ghk}};
% Text Node
\draw (365.2,199.73) node [anchor=north west][inner sep=0.75pt]   [align=left] {{\footnotesize kh}};
% Text Node
\draw (183.6,228.93) node [anchor=north west][inner sep=0.75pt]   [align=left] {{\scriptsize 0}};
% Text Node
\draw (307.2,193.33) node [anchor=north west][inner sep=0.75pt]   [align=left] {{\scriptsize 1}};
% Text Node
\draw (454.8,228.53) node [anchor=north west][inner sep=0.75pt]   [align=left] {{\scriptsize 3}};
% Text Node
\draw (318,25.33) node [anchor=north west][inner sep=0.75pt]   [align=left] {{\scriptsize 2}};
% Text Node
\draw (258.8,173.13) node [anchor=north west][inner sep=0.75pt]    {$z_{2}$};
% Text Node
\draw (323.6,71.93) node [anchor=north west][inner sep=0.75pt]    {$z_{1}$};
% Text Node
\draw (368,152.53) node [anchor=north west][inner sep=0.75pt]    {$z_{3}^{-1}$};
% Text Node
\draw (312.94,216.52) node [anchor=north west][inner sep=0.75pt]    {$z_{1} \ ^{h} z_{2} z_{3}$};
\end{tikzpicture}
    \end{tikzcd}
\end{center}
where the parameters in $\omega$ are chosen according to their simplicial ordering (skipping the \textbf{013} face which is determined by the other faces). The other group elements entering each cocycle in the discrete torsion phase are computed similarly.

\subsubsection{Classification by degree-3 cohomology} \label{ssec:deg3coh}

In the previous section, we outlined the discrete torsion phases in terms of cocycles of the degree-3 cohomology of $\Gamma$. We now explain this cohomology group. Throughout this paper, we will assume trivial actions of $G$ and $K$ on $U(1)$, but will admit nontrivial actions of $G$ on $K$.

The group extension
$$1\longrightarrow BK\longrightarrow \Gamma\longrightarrow G\longrightarrow 1$$
becomes a fibration of topological spaces
\begin{equation}
    B^2K\longrightarrow B\Gamma\longrightarrow BG
\end{equation}
where $B^2K$, $B\Gamma$ and $BG$ are the classifying spaces of $BK$, $\Gamma$, and $G$, respectively. One can compute the cohomology of any total space of a fibration in terms of the base space and the fiber using the Leray-Serre spectral sequence \cite{kochman1996bordism}. In particular, we are interested in computing $H^3_{sing}(B\Gamma,U(1))$, which for simplicity we denote as $H^3(\Gamma,U(1))$. 

To construct the $E_2$ page, we first state the relevant cohomology groups of $H_{sing}(B^2K,U(1))$. Since it is a connected topological space, we have 
\begin{displaymath}
H^0_{sing}(B^2K,U(1))=U(1). 
\end{displaymath}
For $H^1_{sing}(B^2K,U(1))$, it is possible to use the fibration
\begin{equation}
    BK\longrightarrow * \longrightarrow B^2K
\end{equation}
and another Serre sequence to conclude that $H^1_{sing}(B^2K,U(1))=0$. Moreover, in \cite{eilenberg1954groups} it is derived that $H^2_{sing}(B^2K,U(1))=H^1_{sing}(BK,U(1))=H^1(K,U(1))$ and that $H^3_{sing}(B^2K,U(1))=\text{Extabel}(K,U(1))$, where the latter is the group of abelian extensions of $K$ by $U(1)$, in other words, the subgroup of $H^2(K,U(1))$ whose 2-cocycles are symmetric (see e.g.~Proposition 4.8 in \cite{herrero2018extensions}).

We denote $e:=\text{Extabel}(K,U(1))$ for shorthand. The $E_2$ page is
\begin{center}
    \begin{tabular}{l|lllll}
     $e$& $H^0(G,e)$& $H^1(G,e)$& $H^2(G,e)$&$H^3(G,e)$ & $H^4(G,e)$\\
     $\hat{K}$ & $H^0(G,\hat{K})$ & $H^1(G,\hat{K})$ & $H^2(G,\hat{K})$ & $H^3(G,\hat{K})$ & $H^4(G,\hat{K})$ \\
     $0$ & $0$ & $0$ & $0$ & $0$ & $0$ \\
     $U(1)$ & $U(1)$ & $H^1(G,U(1))$ & $H^2(G,U(1))$ & $H^3(G,U(1))$ & $H^4(G,U(1))$\\ \hline
     & $0$ & $1$ & $2$ & $3$ & $4$
\end{tabular}
\end{center}

In this table and below, $H^i(G,\hat{K})$ denotes the group cohomology of $G$ with coefficients in the abelian group $\hat{K}$ equipped with the action of $G$ induced by $\alpha$.  In particular,
note that we wrote $H^0(G,e)$ and $H^0(G,\hat{K})$ and not just $e$ and $\hat{K}$, respectively, since $G$ does carry an action on $K$. Since we are only interested in $H^3(\Gamma,U(1))$, we only need to concentrate on what maps from or to the groups whose coordinates add to $3$, in this case, it is $H^0(G,e)$, $H^1(G,\hat{K})$, $0$, and $H^4(G,U(1))$. The only $d_2$ differential that is relevant in this case is the one that maps $d_2^{0,3}:H^0(G,e)\to H^2(G,\hat{K})$. The relevant groups on page $E_3$ are
\begin{center}
    \begin{tabular}{l|lllll}
     $e$& $\text{ker}(d_2^{0,3})$ & - & - &- & -\\
     $\hat{K}$ & $H^0(G,\hat{K})$ & $H^1(G,\hat{K})$ & - & - & - \\
     $0$ & $0$ & $0$ & $0$ & $0$ & $0$ \\
     $U(1)$ & $U(1)$ & $H^1(G,U(1))$ & $H^2(G,U(1))$ & $H^3(G,U(1))$ & $H^4(G,U(1))$\\ \hline
     & $0$ & $1$ & $2$ & $3$ & $4$
\end{tabular}
\end{center}

The differential $d_3$ is not relevant for $\text{ker}(d_2^{0,3})$ since it lands on a trivial group. It is relevant for $H^3(G,U(1))$ since $d_3^{0,2}: H^0(G,\hat{K})\to H^3(G,U(1))$. As noted in \cite{Yu:2020twi}, the differential $d_3$ has a clear connection with the Postnikov extension class $\beta_{\Gamma}\in H^3(G,K)$. Indeed, $d_3^{0,3}=\ \langle-,\beta_{\Gamma}\rangle$ (that is, precomposition with $\beta$). Meanwhile, $d_3^{1,2}=\ \langle-\cup\beta_{\Gamma}\rangle$. It is also relevant to $H^1(G,\hat{K})$ since $d_3^{1,2}:H^1(G,\hat{K})\to H^4(G,U(1))$. Thus, the relevant groups on page $E_4$ are
\begin{center}
    \begin{tabular}{l|lllll}
     $e$& $\text{ker}(d_2^{0,3})$ & - & - &- & -\\
     $\hat{K}$ & - & $\text{ker}(d_3^{1,2})$ & - & - & - \\
     $0$ & - & - & $0$ & - & - \\
     $U(1)$ & - & - & - & $\text{coker}(d_3^{0,2})$ & $\text{coker}(d_3^{1,2})$\\ \hline
     & $0$ & $1$ & $2$ & $3$ & $4$
\end{tabular}
\end{center}

Finally, the only relevant $d_4$ differential is $d_4^{0,3}:\text{ker}(d^{0,3}_2)\to \text{coker}(d_3^{1,2})$. This will be the last relevant differential, so that the relevant groups in the stabilized page $E_{\infty}$ are
\begin{center}
    \begin{tabular}{l|lllll}
     $e$& $\text{ker}(d_4^{0,3})$ & - & - &- & -\\
     $\hat{K}$ & - & $\text{ker}(d_3^{1,2})$ & - & - & - \\
     $0$ & - & - & $0$ & - & - \\
     $U(1)$ & - & - & - & $\text{coker}(d_3^{0,2})$ & -\\ \hline
     & $0$ & $1$ & $2$ & $3$ & $4$
\end{tabular}
\end{center}

Now that one has the relevant groups, one ends up with an extension problem. In simple terms, there is always a filtration $0=F^3_{-1}\subset F^3_0\subset\cdots F^3_2\subset F^3_3=H^3(\Gamma,U(1))$ that defines short exact sequences 
\begin{equation}
    0\longrightarrow F^3_{i-1}\longrightarrow F^3_i \longrightarrow E^{3-i,i}_{\infty}\longrightarrow 0
\end{equation}
The first two short exact sequences give that $F^3_0 = F^3_1= \text{coker}(d^{0,2}_3)$. One then needs to solve the extension problem for $F^3_2$ in
\begin{equation}
    0\longrightarrow \text{coker}(d^{0,2}_3)\longrightarrow F^3_2\longrightarrow \text{ker}(d_3^{1,2})\longrightarrow 0
\end{equation}
which then determines $H^3(\Gamma,U(1))$ through the extension
\begin{equation}
    0\longrightarrow F^3_2\longrightarrow H^3(\Gamma,U(1))\longrightarrow \text{ker}(d_4^{0,3})\longrightarrow 0.
\end{equation}
The two sequences above summarize the computation of $H^3(\Gamma,U(1))$ for $\Gamma$ a general 2-group.

It becomes difficult to solve these extension problems without further information about the groups involved. One can nevertheless note several properties. First, note that $\text{coker}(d_3^{0,2})$, morally, consists on the 3d discrete torsion choices for $G$ modulo the ones that factor through the Postnikov extension class (with an appropriate notion of $G$-invariance since we are considering $H^0(G,\hat{K})$, not just $\hat{K}$). Moreover, we can follow the extensions and note that, in particular, $\text{coker}(d_3^{0,2}) < H^3(\Gamma,U(1))$. This will play a role in a particular scenario of decomposition.

In passing, we note that there are cases in which the computation simplifies, notably when $H^3_{sing}(B^2K,U(1))=0$, which happens, for example, when $K=\Z_n$, and where the action of $G$ on $K$ is trivial. The special case of $K=\Z_{2n+1}$ is covered in \cite{Yu:2020twi}.

\section{General decomposition conjecture}
\label{sect:decomp-conj}

As has been discussed elsewhere (see e.g.~\cite{Hellerman:2006zs,Sharpe:2022ene,Pantev:2022kpl,Pantev:2022pbf,Robbins:2020msp}), decomposition is the statement that
a local $d$-dimensional quantum field theory with a global $(d-1)$-form symmetry is equivalent to a disjoint union of quantum field theories, known as universes.  In the previous section, we defined orbifolds by 2-groups, meaning, among other things, a gauge theory with a 
gauged trivially-acting 1-form symmetry group, which yields a global 2-form symmetry group, hence a decomposition in a three-dimensional theory.

In this section, we state a precise form for the decomposition conjecture for 3d orbifolds by 2-groups, and discuss connections to other dimensions.
The decomposition described here generalizes that discussed in \cite{Pantev:2022kpl};
for example, we will find examples in which the universes of decomposition are not just copies of one another with different choices of discrete torsion.  Such more general examples are known in two-dimensional theories, but the three-dimensional
examples discussed in \cite{Pantev:2022kpl} were all simpler.

In later sections we will check the decomposition conjecture in a variety of examples.

\subsection{Decomposition in three-dimensional orbifolds}

Given any 2-group $\Gamma$ presented as a crossed module $(\delta: \Gamma_1 \rightarrow \Gamma_0$, $\overline{\alpha}: \Gamma_0 \rightarrow {\rm Aut}(\Gamma_1))$, 
define the group $G$ to be the cokernel of $\delta$ and $K$ its kernel,
so that one has
the short exact sequence
\begin{equation}  \label{eq:longex}
    1 \: \longrightarrow \: K \: \longrightarrow \:
    \Gamma_1 \: \stackrel{\delta}{\longrightarrow} \: \Gamma_0 \: \stackrel{\pi}{\longrightarrow} \: G \:
    \longrightarrow \: 1.
\end{equation}
We conjecture that decomposition takes the form
\begin{equation}   \label{eq:basic-conj}
    {\rm QFT}\left( [X/\Gamma] \right) \: = \:
    {\rm QFT}\left( \left[ \frac{X \times \hat{K} }{G} \right]_{\omega} \right),
\end{equation}
where $\hat{K}$ is the set of irreducible representations of $K$,
and $\omega$ denotes discrete torsion which we will describe momentarily.

The group  $G$ acts on $\hat{K}$ by sending a representation $\psi: K \rightarrow GL(n,{\mathbb C})$ to the
representation $g \cdot \psi$ defined by
\begin{equation}  \label{eq:g-action-hat-k}
    (g \cdot \psi)(z) \: = \: \psi\left( \alpha(g)(z) \right),
\end{equation}

Next, we turn to the description of the discrete torsion $\omega$ on each summand.
Let $\{ \rho_a \}$ be a collection of irreducible representations of $K$ chosen so that the equivalence
classes $[\rho_a]$ are representatives of the orbits of the $G$ action on $\hat{K}$. For each
$\rho_a$, define $H_a \subset G$ to be the stabilizer of $\rho_a$ in $\hat{K}$. The decomposition 
conjecture~(\ref{eq:basic-conj}) can then be written
\begin{equation}\label{eq:3ddecomp}
      {\rm QFT}\left( [X/\Gamma] \right) \: = \:
    {\rm QFT}\left( \coprod_{a}\left[ X/H_a \right]_{\hat{\omega}_a} \right),
\end{equation}

Diagrammatically, Equation~(\ref{eq:3ddecomp}) has a very straightforward interpretation. Recall from Section \ref{sect:basic-defn} that since the $1$-form symmetry $BK$ acts trivially, it gives rise to local operators, also labeled by group elements of $K$. On the gauged theory, we expect a \textit{dual} or \textit{quantum} symmetry, where in particular the corresponding codimension $3$ (point-like) defects are labeled by $\hat{K}$. However, as defined in Equation~(\ref{eq:g-action-hat-k}), $G$ has an action on $\hat{K}$. Taking this action into account, the resulting local operators in the gauged theory, as shown below, are not labeled by $\hat{K}$ but by equivalence classes of representations in $\hat{K}$ described by the orbits of the $G$-action. These local operators are precisely \textit{universe projectors} $\Pi_{\rho_a}$. This is what is stated by the conjecture.
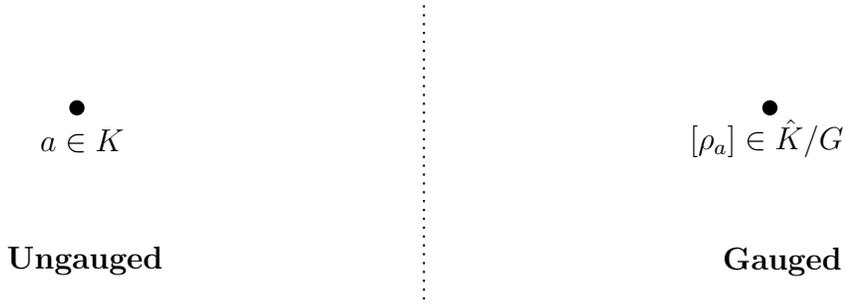
\begin{center}

\tikzset{every picture/.style={line width=0.75pt}} %set default line width to 0.75pt        

\begin{tikzpicture}[x=0.75pt,y=0.75pt,yscale=-1,xscale=1]

%Shape: Circle [id:dp15513497199785475] 
\draw  [fill={rgb, 255:red, 0; green, 0; blue, 0 }  ,fill opacity=1 ] (171.66,151) .. controls (171.66,149.16) and (173.16,147.66) .. (175,147.66) .. controls (176.84,147.66) and (178.34,149.16) .. (178.34,151) .. controls (178.34,152.84) and (176.84,154.34) .. (175,154.34) .. controls (173.16,154.34) and (171.66,152.84) .. (171.66,151) -- cycle ;
%Shape: Circle [id:dp6884434287032462] 
\draw  [fill={rgb, 255:red, 0; green, 0; blue, 0 }  ,fill opacity=1 ] (521.16,151) .. controls (521.16,149.16) and (522.66,147.66) .. (524.5,147.66) .. controls (526.34,147.66) and (527.84,149.16) .. (527.84,151) .. controls (527.84,152.84) and (526.34,154.34) .. (524.5,154.34) .. controls (522.66,154.34) and (521.16,152.84) .. (521.16,151) -- cycle ;
%Straight Lines [id:da665257171890929] 
\draw  [dash pattern={on 0.84pt off 2.51pt}]  (350,99.25) -- (350,250.75) ;

% Text Node
\draw (155,160.4) node [anchor=north west][inner sep=0.75pt]    {$a\in K$};
% Text Node
\draw (482.5,155.4) node [anchor=north west][inner sep=0.75pt]    {$[ \rho _{a}] \in \hat{K} /G$};
% Text Node
\draw (185.17,228) node   [align=left] {\begin{minipage}[lt]{68pt}\setlength\topsep{0pt}
\textbf{Ungauged}
\end{minipage}};
% Text Node
\draw (546.17,228.5) node   [align=left] {\begin{minipage}[lt]{68pt}\setlength\topsep{0pt}
\textbf{Gauged}
\end{minipage}};

\end{tikzpicture}

\end{center}

Now, we make precise the discrete torsion phases $\hat{\omega}_a$. Since $K$ is abelian, the representations are one-dimensional. To define $\omega_a \in H^3(H_a,U(1))$, we will need the extension cocycle $\beta_{\Gamma}$ whose class is in $H^3(G,K)$. We recall how it is constructed following \cite[section 6.6.12]{weibel1995introduction}. 
To that end, let $\sigma: G \rightarrow \Gamma_{0}$ be a section of the projection
$\pi: \Gamma_{0} \rightarrow G$.
For $g_1, g_2 \in G$,
define 
\begin{equation}
    [g_1,g_2] \: = \: \sigma(g_1) \sigma(g_2) \sigma(g_1 g_2)^{-1} \: \in \: \Gamma_{0}.
\end{equation}
It is straightforward to check that $[g_1, g_2] \in \ker \pi$.
Then, it can be shown that
\begin{equation}
    [g_1, g_2] \, [g_1 g_2, g_3] \: = \: \sigma(g_1) [g_2, g_3] \sigma(g_1)^{-1} \,
    [g_1, g_2 g_3].
\end{equation}
Since $[g_1, g_2] \in \ker \pi = {\rm im}\, \delta$, we can lift this expression to a 
relation in $\Gamma_1$. However, after doing so, it will only hold up to an element
of $K$, which will define our cocycle.  In detail, let $[[g_1, g_2]] \in \Gamma_1$ denote
a lift of $[g_1, g_2] \in \ker \pi$, so that $\delta( [[g_1, g_2]] ) = [g_1, g_2]$,
and define
\begin{equation}
    \beta_{\Gamma}(g_1, g_2, g_3) \: = \:
    \sigma(g_1) [[ g_2, g_3]] \sigma(g_1)^{-1} \,
    [[ g_1, g_2 g_3]] \,
    [[ g_1 g_2, g_3]]^{-1} \,
    [[ g_1, g_2]]^{-1},
\end{equation}
then just as in \cite[section 6.6.12]{weibel1995introduction}, it can be shown that
$\beta_{\Gamma}(g_1, g_2, g_3)$ defines a cocycle in $Z^3(G,K)$. and the corresponding element of
$H^3(G,K)$ is independent of choices. Since we have a prescribed $G$-action on $K$ given by $\alpha:G\to\text{Aut}(K)$, the cocycle identity that $\beta_{\Gamma}$ satisfies is
\begin{equation}\label{eq:3gkcocycle}
    \alpha(g_1)(\beta_{\Gamma}(g_2,g_3,g_4))-\beta_{\Gamma}(g_1g_2,g_3,g_4)+\beta_{\Gamma}(g_1,g_2g_3,g_4)-\beta_{\Gamma}(g_1,g_2,g_3g_4)+\beta_{\Gamma}(g_1,g_2,g_3)=0
\end{equation}
Because $H_a< G$, we can restrict this cocycle to $\beta_{\Gamma}^a$ classified by $H^3(H_a,K)$. We now define $\hat{\omega}_a$ as 
\begin{equation}  \label{eq:omega-a}
\hat{\omega}_a = \rho_a\circ \beta^a_{\Gamma} 
\end{equation}

Given that $K$ acts trivially on $U(1)$, $[\rho_a]\in H^1(K,U(1))$ is a homomorphism. Note that $$\rho_a(\alpha(g_1)(\beta_{\Gamma}(g_2,g_3,g_4)))=\tensor[^{g_1}]{\rho}{_a}(\beta_{\Gamma}(g_2,g_3,g_4))=\rho_a(\beta_{\Gamma}(g_2,g_3,g_4))$$ since $g_1\in H_a$ and $H_a$ is the stabilizer of $\rho_a$. Then, composing the cocycle identity~(\ref{eq:3gkcocycle}) with $\rho$ gives the cocycle identity for cocycles in $H^3(H_a,U(1))$ with trivial action of $H_a$ on $U(1)$, that is, a choice of discrete torsion.

We will verify this conjecture in numerous examples in Section \ref{sec:examples}.

Clearly, if the $G$-action on $K$ is trivial, then the decomposition conjecture reduces to
    \begin{equation}\label{eq:3dtrivactdecomp}
      {\rm QFT}\left( [X/\Gamma] \right) \: = \:
    {\rm QFT}\left( \coprod_{\rho}\left[ X/G \right]_{\rho\circ\beta_{\Gamma}} \right),
\end{equation}
as claimed in \cite{Pantev:2022kpl}.

\subsection{Decomposition in the presence of discrete torsion}
\label{sect:decomp-dt}

We briefly comment on the decomposition of 3d 2-group orbifolds with nontrivial choices of discrete torsion. The analogous two-dimensional scenario was explored in \cite{Robbins:2020msp}. As explained in section \ref{ssec:deg3coh}, the discrete torsion choices are classified by $H^3(\Gamma,U(1))$, which is an extension of 
\begin{equation}
    \text{ker}(d_4^{0,3}: \: \text{ker}(d^{0,3}_2) \: \longrightarrow \: \text{coker}(d_3^{1,2}))
\end{equation}
(obtained from two quotients of\footnote{
$\text{Extabel}(K,U(1))$ is a subgroup of $H^2(K,U(1))$ given by the 2-cocycles that determine abelian extensions.
} $H^0(G,\text{Extabel}(K,U(1)))$)
by $F_2^3$ which is itself an extension of 
\begin{equation}
   \text{ker}(d_3^{1,2}: \: H^1(G,\hat{K}) \: \longrightarrow \: H^4(G,U(1))) 
\end{equation}
by
\begin{equation}
    \text{coker}(d_3^{0,2}: \: H^0(G,\hat{K}) \: \longrightarrow \:  H^3(G,U(1))).
\end{equation}

These extensions show that the latter quotient group of $H^3(G,K)$ is a genuine subgroup of $H^3(\Gamma,U(1))$. Thus, we can meaningfully talk about choosing discrete torsion $\omega'$ for $\Gamma$ that only comes from that of $G$. Since these $U(1)$-phases are assigned as $\omega'(g,h,k;z_1,z_2,z_3)=\omega'(\pi(g),\pi(h),\pi(k))$, it is easy to see that such a choice of discrete torsion will give a constant contribution to the decomposition in Equation~(\ref{eq:3ddecomp}). This will not depend upon the particular representative in the quotient of $H^3(G,U(1))$ that embeds into $H^3(\Gamma,U(1))$ because the decomposition precisely sums over all discrete torsion choices in the quotient class. This means that for a discrete torsion choice $[\omega']\in \text{coker}(d_3^{0,2})<H^3(\Gamma,U(1))$, the decomposition becomes
\begin{equation}\label{eq:3ddecompdt}
      {\rm QFT}\left( [X/\Gamma] \right) \: = \:
    {\rm QFT}\left( \coprod_{a}\left[ X/H_a \right]_{\hat{\omega}_a\omega'} \right),
\end{equation}

The other choices of discrete torsion are much more involved to describe. However, the close similarity to the 2d case suggests that a related decomposition as the one described in \cite{Robbins:2020msp} also occurs in 3d, and likely in higher dimensions. This is made explicit by studying the relevant Serre spectral sequence, to which we turn now.

\subsection{Relation to other dimensions}

In the 2d case, given an ordinary group $\Gamma$ with trivially-acting subgroup $K \subset \Gamma$, it was argued
that the quantum field theory of the two-dimensional orbifold obeyed
\begin{equation}
    {\rm QFT}\left( [X/\Gamma] \right) \: = \: {\rm QFT}\left( \left[ \frac{ X \times \hat{K} }{G} \right]_{\hat{\omega}} \right),
\end{equation}
where $G = \Gamma/K$, and $\hat{\omega}$ was defined similarly, using the central extension class $\beta\in H^2(G,K)$. 

The root of the similarity of these two decomposition stories partly comes from the Serre spectral sequence determined by the extension class. Recall from Section \ref{ssec:deg3coh} that the contribution of $G$ alone to the degree-3 cohomology of $\Gamma$ is through the quotient 
\begin{equation}\label{eq:3dquot}
    H^3(G,U(1))/(\text{Im}(-\circ\beta_{\Gamma}:H^0(G,H^1(K,U(1)))\to H^3(G,U(1))) \subset H^3(\Gamma,U(1)).
\end{equation}
The decomposition in (\ref{eq:3dtrivactdecomp}) for trivial $G$-action on $K$ is exactly summing over all the discrete torsion choices on $G$ that can be expressed as composition of representations of $K$ with the extension cocycle, which are the choices that are left out above. This is more clearly understood in terms of the projection of $\Gamma$-twisted sectors to $G$-twisted sectors, as explained in Section \ref{eq:3dtrivactdecomp}. The more general decomposition (\ref{eq:3ddecomp}) is understood in a similar way. The main difference is that because of the additional condition $z_1\tensor[^h]{z}{_2}z^3=\tensor[^g]{z}{_3}z_2\tensor[^k]{z}{_1}$ for 2-orbifolds, the $\Gamma$-twisted sectors do not cover all twisted sectors of $G$ with the same multiplicity. This causes the assembly of orbifolds by subgroups of $G$ determined by how these stabilize the representations left out in (\ref{eq:3dquot}). Moreover, we sum over discrete torsion choices that come from the composition of the extension cocycle and \textit{fixed} representations, as suggested by (\ref{eq:3dquot}), since the quotient if by choices coming from $H^0(G,H^1(K,U(1)))$.

In two dimensions, essentially the same phenomenon happens. Suppose we have an extension
\begin{equation}
    1\longrightarrow K\longrightarrow\Gamma\longrightarrow G\longrightarrow 1
\end{equation}
classified by $\beta\in H^2(G,K)$ with a $G$-action on $K$. We can compute $H^2(\Gamma,U(1))$ using the Serre sequence. The relevant elements in the $E_2$ page are
\begin{center}
    \begin{tabular}{l|lllll}
     $H^2(K,U(1))$ & $H^0(G,H^2(K,U(1)))$ & - & - & - & - \\
     $H^1(K,U(1))$ & $H^0(G,H^1(K,U(1)))$ & $H^1(G,H^1(K,U(1)))$ & - & - & - \\
     $U(1)$ & - & - & $H^2(G,U(1))$ & $H^3(G,U(1))$ & -\\ \hline
     & $0$ & $1$ & $2$ & $3$ & $4$
\end{tabular}
\end{center}
Note that there is only one nontrivial differential acting on $H^2(G,U(1))$, which is
\begin{equation}
    d_2^{0,2}=-\circ\beta:H^0(G,H^1(K,U(1))\to H^2(G,U(1))
\end{equation}
which is exactly the same setting as in the 3d case, that is, a differential given by composition with the extension class. Also note that $H^2(\Gamma,U(1))$ is again an extension of a quotient of $H^d(G,U(1))$ (where here $d=2$), a quotient of $H^1(G,H^1(K,U(1)))$, and two quotients of $H^0(G,H^2(K,U(1)))$. The only difference is that here we have $H^2(K,U(1))$, whereas in 3d and in higher dimensions we will have the stabilization of this, that is, $\text{Extabel}(K,U(1))$. Since the discrete torsion here has heuristically the same form as the 3d case, this seems to suggest that a similar decomposition for nontrivial discrete torsion will occur in 3d.

There is, however, an important difference. While in the 3d case, as explained in Section \ref{ssec:central}, at least for a trivial $G$-action on $K$, all the $G$-twisted sectors are equally covered by $\Gamma$-twisted sectors, this is not true in 2d. The $G$-twisted sectors not covered by $\Gamma$-twisted sectors are exactly the ones eliminated by summing over discrete torsion choices coming from composing with representations of $K$, though. 

This is exemplified in the $D_4$ orbifold example described in \cite{Pantev:2005rh},
where $D_4$ is given by the extension
\begin{equation}
    1\longrightarrow \Z_2\longrightarrow D_4\longrightarrow \klein\longrightarrow 1
\end{equation}
and ${\mathbb Z}_2 \subset D_4$ acts trivially.
The $Z(a,b)$ $\klein$-twisted sector does not lift to a $D_4$-twisted sector, but these sectors are eliminated by summing over the described discrete torsion choices, as $\beta(a,b)$ gives a nontrivial value. This, however, does not work in a three-dimensional $D_4$ orbifold. For instance, the 3d $\klein$-twisted sector $Z(1,a,b)$ does not lift to a sector in $D_4$, yet it cannot be eliminated through discrete torsion since the cocycles are normalized. In a sense, orbifolds by extensions of 1-groups are about lifting information of commutativity from $G$ to $\Gamma$. This cannot be done in general, but only in 2d can we eliminate the sectors whose lifts do not commute using discrete torsion so that we get a decomposition. In the 3d case, we are trying to lift commutativity information from $G$ to $\Gamma_0$ for $\Gamma$ a 2-group. But by exactness, we will always be able to make the lifts commute, up to an image of $\Gamma_1$. There are associativity constraints on the elements of $\Gamma_1$ that realize such commutativity identities in $\Gamma_0$, but since we are not lifting associativity information from $G$, then the constraints are satisfied as shown in Section \ref{ssec:central}. If, on the other hand, $G$ was itself a 2-group, then we would need to lift associativity information, and thus in analogy to the 2d case of 1-groups, we would expect to have twisted sectors in $G$ that do not lift to $\Gamma$ but that only in 3d can be eliminated by summing over discrete torsion.

It is not difficult to generalize these observations to $d$-dimensional theories. Indeed, consider $G$ a group, $K$ an abelian group with a $G$-action, and an extension class $\beta\in H^d(G,K)$. This information determines a \textit{strict} $(d-1)$-group $\Gamma$ (which for $d>3$ does not cover all $(d-1)$-groups anymore). The discrete torsion choices are classified by $H^d(\Gamma,U(1))$, which can be computed using the Serre sequence associated to the fibration
\begin{equation}
    B^{d-1}K\longrightarrow B\Gamma\longrightarrow BG.
\end{equation}
One can show that $H^j(B^{d-1}K,U(1))=0$ for $0<j<d-1$ and $H^{d-1}(B^{d-1}K,U(1))=H^1_{grp}(K,U(1))$. The relevant groups for this discussion in the $E_2$ page are
\begin{center}
    \begin{tabular}{l|lllll}
   $e$ & $H^0(G,e)$ & - & $\cdots$ & - & -\\
   $H^1(K,U(1))$ & $H^0(G,H^1(K,U(1)))$ & - & $\cdots$ & - & - \\
   $\cdots$ & $\cdots$& $\cdots$ & $\cdots$ &$\cdots$ & $\cdots$\\
   $0$ & $0$ & $0$ & $\cdots$ & $0$ & $0$ \\
   $U(1)$ & - & - & $\cdots$& $H^{d}(G,U(1))$& $H^{d+1}(G,U(1))$\\ \hline
   & $0$ & $1$ & $\cdots$ & $d$ & $d+1$
\end{tabular}
\end{center}
where we used that $H^d(B^{d-1}K,U(1))=H^3(B^2K,U(1))=\text{Extabel}(K,U(1))=e$ is a stable result \cite{eilenberg1954groups}.

Note that the only nontrivial differential that acts on $H^d(G,U(1))$ is $$d^{0,d-1}_d=-\circ\beta: H^0(G,H^1(K,U(1)))\to H^d(G,U(1))$$
so that the $G$ only stabilized contribution will be
\begin{equation}
    H^d(G,U(1))/\text{Im}(-\circ\beta)\subset H^d(\Gamma,U(1))
\end{equation}
so that again we are quotienting out the discrete torsion choices that come from composing the extension cocycle with representations of $K$. Since $G$ only carries commutativity information, we expect to be able to lift every $G$-twisted sector to $\Gamma$-twisted sectors as in the 3d case. This is the same setup as in the previous cases.

All in all, we conjecture that for a $d$-dim'l theory by a $(d-1)$-group $\Gamma$ specified by a finite group $G$, a finite abelian group $K$ with a $G$-action, and an extension class $\beta_{\Gamma}\in H^d(G,K)$, there will be a decomposition
\begin{equation}
      {\rm QFT}\left( [X/\Gamma] \right) \: = \:
    {\rm QFT}\left( \coprod_{a}\left[ X/H_a \right]_{\hat{\omega}_a} \right),
\end{equation}
as in Equation~(\ref{eq:3ddecomp}), with $\hat{\omega}_a=\rho_a\circ \beta_{\Gamma}$ and $\rho_a$ a representation of $K$ whose stabilizer is $H_a\subset G$. Moreover, for nontrivial discrete torsion choices for $\Gamma$, it seems reasonable that we will have a similar decomposition as the cases listed in \cite{Robbins:2020msp}, given that $H^n(\Gamma,U(1))$ is always given as an extension in terms of a quotient of $H^d(G,U(1))$, a quotient of $H^1(G,H^1(K,U(1)))$, and two quotients of $H^0(G,e)$, where most of the relevant differentials are given by contraction with the extension class. The only difference would be that not all of $H^2(K,U(1))$ will appear but only $\text{Extabel}(K,U(1))\subset H^2(K,U(1))$, its stabilization. 

Finally, aside from the general remarks we have made,
we leave a detailed study of the decomposition of orbifolds
$[X/\Gamma]$ with discrete torsion for future work.

\section{Examples}\label{sec:examples}

\subsection{Ordinary group}\label{ssec:1grp}

As a basic consistency test, in this section we will consider the case that the
2-group is (weakly equivalent to) an ordinary group, and verify that one recovers
an ordinary group orbifold as a special case of our prescription.

The most general 2-group $\Gamma=(\Gamma_1\xrightarrow{\delta}\Gamma_0)$ that is weakly equivalent to an ordinary group is that for which the boundary homomorphism $\delta$ is injective. In such a case, it follows from the axioms in appendix~\ref{app:background} that $\Gamma_1$ is a normal subgroup of $\Gamma_0$, and that $\Gamma_0$ acts on $\Gamma_1$ by conjugation. This 2-group is weakly equivalent to $\cokerd$. This trivially fits in the short exact sequence
\begin{equation*}
    1 \: \longrightarrow \: 1 \: \longrightarrow \:
    \Gamma_1 \: \hookrightarrow \: \Gamma_0 \:  \stackrel{\pi}{\longrightarrow} \: \cokerd \: \longrightarrow \: 1,
\end{equation*}
where $\pi:\Gamma_0\to\cokerd$ is surjective.

In such a case, in terms of the decomposition conjecture~(\ref{eq:basic-conj}),
$G = {\rm coker} \, \delta$, $K = 1$, with a trivial $G$-action for obvious reasons, and so the decomposition conjecture reduces to the statement
that the $\Gamma$ orbifold should coincide with an ordinary $G$ orbifold:
\begin{equation}
    {\rm QFT}\left( [X/\Gamma] \right) \: = \:
    {\rm QFT}\left( [X/G] \right).
\end{equation}

Next, let us compute partition functions on $T^3$, to check that physics also sees 
such a 2-group orbifold as equivalent to an ordinary group orbifold.

To that end, note that if $Z(g,h,k;z_1,z_2,z_3)$ is a twisted sector in the $[X/\Gamma]$ theory, 
then $Z(\pi(g),\pi(h),\pi(k))$ is a twisted section in $[X/\cokerd]$. This is because $g,h,k\in\Gamma_0$ commute up to an element of $\Gamma_1$, which is the kernel of $\pi$. Now we show that every twisted sector $Z(a,b,c)$ in $[X/\cokerd]$ lifts to a twisted sector in $[X/\Gamma]$. To see this, let $\sigma_1,\sigma_2,\sigma_3:\cokerd\to\Gamma_0$ be independent sections of $\pi$. Since it is assumed that $a,b,c\in\cokerd$ pairwise commute, then their lifts $\sigma_1(a),\sigma_2(b),\sigma_3(c)\in\Gamma_0$ pairwise commute up to elements $z_1,z_2,z_3\in\text{ker}(\pi)=\Gamma_1$. That is, we have 
\begin{eqnarray*}
    \sigma_1(a)\sigma_2(b) &=& \delta(z_1)\sigma_2(b)\sigma_1(a), \\
    \sigma_1(a)\sigma_3(c) &=& \delta(z_2)\sigma_3(c)\sigma_1(a), \\
    \sigma_2(b)\sigma_3(c) &=& \delta(z_3)\sigma_3(c)\sigma_2(b).
\end{eqnarray*}

Recalling that the action of $\Gamma_0$ on $\Gamma_1$ is by conjugation, it is a matter of plugging these equations into the group elements $z_1\tensor[^{\sigma_2(b)}]{z}{_2}z_3$ and $\tensor[^{\sigma_1(a)}]{z}{_3}z_2\tensor[^{\sigma_3(c)}]{z}{_1}$ to see that
\begin{equation*}
    z_1\tensor[^{\sigma_2(b)}]{z}{_2}z_3 = \tensor[^{\sigma_1(a)}]{z}{_3}z_2\tensor[^{\sigma_3(c)}]{z}{_1},
\end{equation*}
so that $Z(\sigma_1(a),\sigma_2(b),\sigma_3(c);z_1,z_2,z_3)$ is a twisted sector in $[X/\Gamma]$. Moreover, since the sections were chosen independently, then every $Z(a,b,c)$ twisted sector in $[X/\cokerd]$ is covered $|\Gamma_1|^3$ times by twisted sectors in $[X/\Gamma]$. Hence, the partition function becomes
\begin{eqnarray}
    Z_{T^3}([X/\Gamma])
    & = &
    \frac{1}{|\Gamma_1|^2|\Gamma_0|}\sum_{(g,h,k;z_1,z_2,z_3)\in\Gamma} Z(g,h,k;z_1,z_2,z_3) ,
    \\
    & = &
    \frac{1}{|\Gamma_1|^2|\Gamma_0|} |\Gamma_1|^3\sum_{(a,b,c)} Z(a,b,c),
    \\
    & = &
    \frac{1}{|\cokerd|}\sum_{(a,b,c)\in\cokerd} Z(a,b,c),
    \\
    & = & Z_{T^3}([X/\cokerd]),
\end{eqnarray}
so that our prescription does reproduce the ordinary orbifold cases whenever the 2-group $\Gamma$ is weakly equivalent to an ordinary group.

This description not only recovers the usual orbifold case, but also reproduces discrete torsion for orbifolds. Recall that for a 2-group $\xmod$, its discrete torsion values are classified by cocycles in $H^3(\Gamma,U(1))$, which is computed using the Serre spectral sequence as described in Section \ref{ssec:deg3coh}. In the special case where $\Gamma$ is weakly equivalent to the 1-group $\cokerd$, it is straightforward to see that $H^3(\Gamma,U(1))\cong H^3(\cokerd,U(1))$. Clearly, since the cohomology only depends on $\cokerd$, then the discrete torsion phases $\omega(g,h,k;z_1,z_2,z_3)$ do not depend on the values of $z_1,z_2,z_3$ nor any of the information coming from $\Gamma_1$, so that $\omega(g,h,h;z_1,z_2,z_3)=\omega(\pi(g),\pi(h),\pi(k))$ where $\pi:\Gamma_0\to\cokerd$ is the quotient map. Hence the partition function for a nontrivial choice of discrete torsion $[\omega]\in H^3(\Gamma,U(1))\cong H^3(\cokerd,U(1))$ with associated phases $\epsilon_{\omega}$ is
\begin{eqnarray}
    Z_{T^3}([X/\Gamma])
    & = &
    \frac{1}{|\Gamma_1|^2|\Gamma_0|}\sum_{(g,h,k;z_1,z_2,z_3)\in\Gamma} \epsilon_{\omega}(\pi(g),\pi(h),\pi(k))Z(g,h,k;z_1,z_2,z_3),
    \\
    & = &
    \frac{1}{|\Gamma_1|^2|\Gamma_0|} |\Gamma_1|^3\sum_{(a,b,c)} \epsilon_{\omega}(a,b,c) Z(a,b,c),
    \\
    & = & \frac{1}{|\cokerd|}\sum_{(a,b,c)\in\cokerd} \epsilon_{\omega}(a,b,c) Z(a,b,c),
    \\
    & = & Z_{T^3}([X/\cokerd]_{\omega}),
\end{eqnarray}
so that we reproduce any orbifold with any of the possible choices of discrete torsion.

\subsection{Central extensions}\label{ssec:central}

Next, we consider the case of a 2-group corresponding to a `central' extension
\begin{equation}
    1 \: \longrightarrow \: BK \: \longrightarrow \: \Gamma \: \longrightarrow \: G \: \longrightarrow \: 1,
\end{equation}
as studied in \cite{Pantev:2022kpl}. 
A central extension is defined by the property that the map
$\alpha: G \rightarrow {\rm Aut}(K)$ is trivial, which distinguishes it from more general 2-groups.
The reader should note that the extension class $[\beta_{\Gamma}]\in H^3(G,K)$ is allowed to be nontrivial.

Let us begin by specializing our general decomposition
conjecture~(\ref{eq:basic-conj}) to this case.  The general statement
is 
\begin{equation}  
    {\rm QFT}\left( [X/\Gamma] \right) \: = \:
    {\rm QFT}\left( \left[ \frac{X \times \hat{K} }{G} \right]_{\omega} \right).
\end{equation}
In this case, however, since $\alpha$ is trivial, the action of
$G$ on $\hat{K}$ is trivial, as can be seen immediately 
from~(\ref{eq:g-action-hat-k}).  As a result, the general statement
reduces to
\begin{equation}  
    {\rm QFT}\left( [X/\Gamma] \right) \: = \:
    {\rm QFT}\left( \left[ \frac{X \times \hat{K} }{G} \right]_{\omega} \right) \: = \:
    {\rm QFT}\left( \coprod_{ \rho \in \hat{K} } [X/G]_{\omega(\rho)}. \right),
\end{equation}
which is of the same form as that in \cite{Pantev:2022kpl}.
Finally, we need to compute the discrete torsion choices $\omega(\rho)$.
Following the analysis of section~\ref{sect:decomp-conj}, since $G$ acts trivially on $\hat{K}$, each isomorphism class in $\hat{K}$ is a $G$-orbit, so the stabilizers are all of $G$, i.e.~$H_a = G$ for all $a$.  Then, equation~(\ref{eq:omega-a}) reduces to
\begin{equation}
    \omega(\rho) \: = \: \rho \circ \beta_{\Gamma},
\end{equation}
which reproduces the discrete torsion on universes in \cite{Pantev:2022kpl}.
Altogether, we see that in this case ($\alpha$ trivial),
our general decomposition conjecture~(\ref{eq:basic-conj}) reduces to that given and checked in \cite{Pantev:2022kpl} for the case of
$\alpha$ trivial (defining central extensions).

Next, we compute partition functions on $T^3$, to check the prediction above.  Of course, partition functions were also computed in \cite{Pantev:2022kpl}; here, we refer to the specialization of our computations for more general
groupoids, as a thorough consistency check.

To that end, first note that our partition function expression~(\ref{eq:Z:invt}) reduces to the expression for $T^3$ partition functions in
\cite[equ'n (4.18)]{Pantev:2022kpl}.  It is already in nearly the same form;
the difference lies in the sum over tuples.
In the general case, the sum is over tuples $(g,h,k;k_1,k_2,k_3)$ such that $g,h,k\in G$ pairwise commute, satisfy $\beta_{\Gamma}(g,h,k)=1\in K$, and $k_1, k_2, k_3\in K$ satisfy $k_1\tensor[^h]{k}{_2}k_3=\tensor[^g]{k}{_3}k_2\tensor[^k]{k}{_1}$.  In the partition functions in \cite{Pantev:2022kpl}, the last condition was omitted.
This is consistent because
in the case $\alpha$ is trivial,
the last condition becomes $k_1 k_2 k_3 = k_3 k_2 k_1$, which is trivial
as $K$ abelian.  Thus, our expression for partition functions correctly
specializes to that of \cite[equ'n (4.18)]{Pantev:2022kpl} in the case of central extensions.

Now, for completeness, let us review how these partition functions are consistent with
decomposition.  (The reader should note that an equivalent verification also appeared in
\cite{Pantev:2022kpl}, though we will describe it differently.)

To that end, we first compare the twisted sectors of $[X/\Gamma]$ with those of $[X/G]$. If $Z(g,h,k;z_1,z_2,z_3)$ is a twisted sector in $[X/\Gamma]$, then $Z(\pi(g),\pi(h),\pi(k))$ is a twisted sector in $[X/G]$, since $g,h,k$ pairwise commute in $\Gamma_0$ up to the images of $z_1,z_2,z_3$ under $\delta$, but $\pi\circ\delta=1$ by exactness of the sequence. Moreover, if $Z(g,h,k)$ is a twisted sector in $[X/G]$ then it lifts to a twisted sector $Z(g',h',k';z_1,z_2,z_3)$. The argument is almost identical to that in Section \ref{ssec:1grp}. Since $g,h,k$ pairwise commute in $G$, then their (independent) lifts $g',h',k'$ commute in $\Gamma_0$ up to elements $\delta(z_1),\delta(z_2),\delta(z_3)\in\text{Im}(\delta)$. Moreover, they satisfy the equation $z_1\tensor[^{h'}]{z}{_2}z_3=\tensor[^{g'}]{z}{_3}z_2\tensor[^{k'}]{z}{_1}$. This is because $G$ acts trivially on $K$, so that by the crossed module identities the only action that $\Gamma_0$ can have on $\Gamma_1$ is by conjugation whenever it is defined. Finally, we note the multiplicities of such lifts. Since $g,h,k$ can be lifted independently, then this gives a multiplicity of $|\text{Im}(\delta)|^3=\frac{|\Gamma_1|^3}{|K|^3}$. Moreover, since $K\hookrightarrow Z(\Gamma_1)$, each $z$ has a $|K|$-fold cover.
Thus, in summary, the twisted sectors in $[X/\Gamma]$ always descend to twisted sectors in $[X/G]$ and, moreover, they cover each one with a multiplicity of $\frac{|\Gamma_1|^3}{|K|^3}\cdot |K|^3 = |\Gamma_1|^3$.

We also need to understand $G$ discrete torsion on the universes $[X/G]$.  We will describe them next
more formally than was done in \cite{Pantev:2022kpl}.
To that end, first note that the projection map $\Pi:\Gamma\to G$ defined in Section~\ref{ssec:trivially-acting} induces a map in cohomology $\Pi^*:H^3(G,U(1))\to H^3(\Gamma,U(1))$. Now, we know that 
\begin{equation}
    \text{coker}\left(-\circ \beta_{\Gamma}:H^1(K,U(1))\to H^3(G,U(1))\right) \: < \: H^3(\Gamma,U(1)).
\end{equation}
Note that we have replaced $H^0(G,H^1(K,U(1)))$ with $H^1(K,U(1))$ since $G$ acts trivially on $K$. Moreover, $\Pi^*$ is exactly the quotient map, since it comes from the short exact sequence of $\Gamma$. Thus, we have that
\begin{equation}
    \Pi^*:H^3(G,U(1))\to \text{coker}\left(-\circ \beta_{\Gamma}:H^1(K,U(1))\to H^3(G,U(1))\right)
\end{equation}
Now, our $\Gamma$ orbifolds have, by assumption, trivial 2-group discrete torsion.
The reader should note the fiber of $\Pi^*$ above over the trivial element of $H^3(\Gamma,U(1))$
is nontrivial in $H^3(G,U(1))$, and, formally, the different universes should sum over choices
of discrete torsion in that fiber, $\text{ker}(\Pi^*)$. Now, since $\Pi^*$ is the quotient map, this says that we must sum over all choices of discrete torsion of $[X/\Gamma]$ that equal $\rho\circ\beta_{\Gamma}$ for $\rho\in H^1(K,U(1))$.

We now apply these observations to the partition function
\begin{eqnarray}
    Z_{T^3}([X/\Gamma])
    & = &
    \frac{1}{|\Gamma_1|^2|\Gamma_0|}\sum_{(g,h,k;z_1,z_2,z_3)\in\Gamma} \epsilon_0(g,h,k;z_1,z_2,z_3) Z(g,h,k;z_1,z_2,z_3) ,
    \\
    & = &
    \frac{1}{|\Gamma_1|^2|\Gamma_0|} |\Gamma_1|^3 \sum_{(g,h,k)\in G} ((\Pi^*)^{-1}(\epsilon_0))(g,h,k) Z(g,h,k),
    \\
    & = &
    \frac{|\Gamma_1|}{|\Gamma_0|}\frac{1}{|H^1(K,U(1))|}\sum_{\rho\in H^1(K,U(1))}\sum_{(g,h,k)\in G} (\rho\circ\beta_{\Gamma})(g,h,k) Z(g,h,k),
    \\
    & = &
    \sum_{\rho\in H^1(K,U(1))} \frac{|\Gamma_1/K|}{|\Gamma_0|}\sum_{(g,h,k)\in G} (\rho\circ\beta_{\Gamma})(g,h,k) Z(g,h,k),
    \\
    & = &
    \sum_{\rho\in H^1(K,U(1))} Z_{T^3}([X/G]_{\rho\circ\beta_{\Gamma}}),
\end{eqnarray}
where $\epsilon_0$ means the discrete torsion phase associated with the trivial choice of discrete torsion. This agrees with \cite[equ'n (4.22)]{Pantev:2022kpl}, and with the conjecture in Eq.~(\ref{eq:3dtrivactdecomp}).

In passing, we emphasize that 2-groups with trivial $\alpha: G \rightarrow {\rm Aut}(K)$ can arise from crossed module presentations with nontrivial $\overline{\alpha}$.  Consider for example the case
$\Gamma_1 = {\mathbb Z}_2 \times {\mathbb Z}_2$, $\Gamma_0 = {\mathbb Z}_4$,
$\delta: \Gamma_1 \rightarrow \Gamma_0$ given by $\delta(a)=\delta(b)=z^2$,
for $\Gamma_1 = {\mathbb Z}_2 \times {\mathbb Z}_2 = \langle a \rangle \times \langle b \rangle$,
and $\overline{\alpha}: \Gamma_0 \rightarrow {\rm Aut}(\Gamma_1)$ given by
$\overline{\alpha}(z)(a)=b$, $\overline{\alpha}(z)(b)=a$.  In this example,
the action of $\cokerd=\Z_2$ on $\ker \delta$ is trivial, so although $\overline{\alpha}$ is nontrivial,
$\alpha: G = {\rm coker}\, \delta \rightarrow {\rm Aut}(\ker \delta)$ is trivial,
hence this crossed module is an example of a central extension.

\subsection{$\Gamma=(\Z_2\times\Z_2\xrightarrow{0} \Z_2)$}

We now consider a crossed module $\Gamma$ (partially) defined by $\Gamma_1 = {\mathbb Z}_2 \times {\mathbb Z}_2$, $\Gamma_0 = {\mathbb Z}_2$, $\delta = 0$, so that $G = {\mathbb Z}_2$ and $K = {\mathbb Z}_2 \times {\mathbb Z}_2$.
We will consider two examples of actions $\alpha: G \rightarrow {\rm Aut}(K)$.

Let us first consider the case that $\alpha$ acts trivially, mapping all of ${\mathbb Z}_2$
to the identity automorphism.  This is an example of a central extension, previously
studied in section~\ref{ssec:central}.  We include the analysis of this special case here for completeness.

In this case, the decomposition conjecture~(\ref{eq:basic-conj})
predicts
\begin{eqnarray}
    {\rm QFT}\left( [X/\Gamma] \right) 
    & = &
    {\rm QFT}\left( \left[ \frac{ X \times \hat{K} }{G} \right] \right),
    \\
    & = &
    {\rm QFT}\left( \coprod_{ \rho \in \hat{K} } [X/G]_{\omega(\rho)} \right),
\end{eqnarray}
where $G = {\mathbb Z}_2$, $K = {\mathbb Z}_2 \times {\mathbb Z}_2$.
The discrete torsion $\omega(\rho) = \rho \circ \beta$ for $\beta \in H^3(G,K)$ classifying the 2-group.  In the present case, $\beta = 0$, hence $\omega(\rho) = 0$ for all $\rho \in \hat{K}$, hence decomposition predicts
\begin{equation}
    {\rm QFT}\left( [X/\Gamma] \right) \: = \: 
    {\rm QFT}\left( \coprod_{\hat{K}} [X/{\mathbb Z}_2] \right).
\end{equation}

The partition function in this case is
\begin{eqnarray*}
    Z_{T^3}(\Z_2\times\Z_2\xrightarrow{0}\Z_2) 
    & = &
    \frac{1}{|\Z_2\times\Z_2|^2|\Z_2|} \sum_{a,b,c\in \Z_2\times\Z_2} \sum_{g,h,k\in \Z_2} Z(g,h,k;a,b,c),
    \\
    & = &
    \frac{|\Z_2\times\Z_2|^3}{|\Z_2\times\Z_2|^2|\Z_2|}|\Z_2|Z_{T^3}(\Z_2),
    \\
    & = &
    |\Z_2\times\Z_2| Z_{T^3}(\Z_2).
\end{eqnarray*}
matching the decomposition prediction above.

Next, we turn to the case $\alpha$ is nontrivial, given by
$\alpha(z)(a)=b$ and 
$\alpha(z)(b)=a$, where we have written ${\mathbb Z}_2 \times {\mathbb Z}_2 =
\langle a, b \rangle$.
Here,
$G = {\mathbb Z}_2$, $K = {\mathbb Z}_2 \times {\mathbb Z}_2$.  $G$ leaves invariant the
trivial irreducible representation of $K$, as well as the irreducible representation that is
nontrivial on both ${\mathbb Z}_2$ factors, but exchanges the other two irreducible representations.
As a result, the decomposition conjecture~(\ref{eq:basic-conj}) implies
\begin{eqnarray}
{\rm QFT}\left( [X/\Gamma] \right)
& = &
{\rm QFT}\left( \left[ \frac{X \times \hat{K} }{G} \right] \right),
\\
& = & 
{\rm QFT}\left( [ X/ {\mathbb Z}_2 ] \, \coprod \, [X/{\mathbb Z}_2] \, \coprod \, \left[ \frac{ X \times {\mathbb Z}_2 }{
{\mathbb Z}_2
}\right] \right),
\\
& = & {\rm QFT}\left( [ X/ {\mathbb Z}_2 ] \, \coprod \, [X/{\mathbb Z}_2] \, \coprod \, X \right).
\end{eqnarray}
Since $\beta_{\Gamma} = 0$, there is no discrete torsion on any universe.

Now, let us compare this prediction to physics.
The partition function in this case is
\begin{eqnarray*}
    Z_{T^3}(\Z_2\times\Z_2\xrightarrow{0}\Z_2)
    & = &
    \frac{|\Z_2\times\Z_2|^2}{|\Z_2\times\Z_2|^2|\Z_2|} %|\Z_2\times\Z_2|^2 
    \bigl(|\Z_2\times\Z_2| Z(1,1,1) 
    \nonumber \\
    & & \hspace*{1in}
    \: + \:|\Z_2| [Z(1,1,z) +Z(1,z,z) + \text{perms.}
    \nonumber \\
    & & \hspace*{1.75in}
    +Z(z,z,z) ]\bigr),
    \\
    & = &|\Z_2| Z_{T^3}([X/\Z_2]) + Z_{T^3}(X),
\end{eqnarray*}
matching the prediction of the decomposition conjecture for this choice of $\alpha: G \rightarrow {\rm Aut}(K)$, but distinct from the result for the previous choice of $\alpha$.

To summarize, the examples of this section explicitly illustrate the role of the action $\alpha$: we have two examples with the same $G$, $K$, and vanishing $\beta$, but different $\alpha: G \rightarrow {\rm Aut}(K)$.
To specify a 2-group $\Gamma$ one must specify all four of $G$, $K$, $\beta$, $\alpha$, and here we see that different choices of $\alpha$ can lead to different results for physical theories.

\subsection{$\Gamma=(\Z_3\xrightarrow{0}\Z_2)$}

We now consider the crossed module $\Gamma = (\Z_3\xrightarrow{0}\Z_2)$ with two different actions $\alpha$, much as we did in the previous example. 

The first case we consider is that of trivial action $\alpha$ of $\Z_2$ on $\Z_3$. 
We begin with the decomposition prediction~(\ref{eq:basic-conj}) for this
case.  In the notation of (\ref{eq:basic-conj}), $G = {\mathbb Z}_2$, $K = {\mathbb Z}_3$,
and since the action $\alpha$ of ${\mathbb Z}_2$ on ${\mathbb Z}_3$ is trivial,
all of the irreducible representations of $K$ are invariant under $G$.

In this case, the decomposition conjecture~(\ref{eq:basic-conj}) predicts that
\begin{eqnarray}
    {\rm QFT}\left( [X/\Gamma] \right)
    & = &
    {\rm QFT}\left( \left[ \frac{X \times \hat{K}}{G} \right] \right),
    \\
    & = & {\rm QFT}\left( \coprod_3 [X/{\mathbb Z}_2] \right).
\end{eqnarray}

The partition function is
\begin{eqnarray*}
    Z_{T^3}(\Z_3\xrightarrow{0}\Z_2) 
    & = & \frac{1}{|\Z_3|^2|\Z_2|}\sum_{a,b,c\in\Z_3} \sum_{g,h,k\in\Z_2} Z(g,h,k;a,b,c),
    \\
    & = & \frac{|\Z_3|^3}{|\Z_3|^2 |\Z_2|} \cdot |\Z_2| Z_{T^3}(X/\Z_2), 
    \\
    & = & 
    |\Z_3| \ Z_{T^3}([X/\Z_2]),
\end{eqnarray*}
agreeing with the prediction of the decomposition conjecture.

Next, we consider a nontrivial $\alpha$, specifically,
the case that $G = \Z_2$ acts on $K =\Z_3$ as $\alpha(z)(a)=a^2$,
where $G = \langle z \rangle$ and $K = \langle a \rangle$.
Here, one of the irreducible representations of $K$ is invariant, but the other two are
interchanged by the action of ${\mathbb Z}_2$.  The decomposition conjecture then predicts
\begin{eqnarray}
    {\rm QFT}\left( [X/\Gamma] \right) & = &
    {\rm QFT}\left( \left[ \frac{ X \times \hat{K}}{G} \right] \right),
    \\
    & = &
    {\rm QFT}\left( [X/{\mathbb Z}_2] \, \coprod \, \left[ \frac{X \times {\mathbb Z}_2}{{\mathbb Z}_2}\right] \right),
    \\
    & = &
     {\rm QFT}\left( [X/{\mathbb Z}_2] \, \coprod \, X \right),
\end{eqnarray}
where the $[X/{\mathbb Z}_2]$ is associated to the trivial representation,
and the $X$ is associated to the two nontrivial representations of $K$.

Now, let us compare to physics.
We have the partition function
\begin{eqnarray*}
    Z_{T^3}(\Z_3\xrightarrow{0}\Z_2) 
    & = &
    \frac{1}{|\Z_3|^2|\Z_2|} \bigl(|\Z_3|^2 Z(1,1,1) 
    \nonumber \\
    & & \hspace*{1in}
    + |\Z_3|^2 [ \left[Z(1,1,z)+Z(1,z,z) +\text{perms.}+Z(z,z,z)\right]\bigr),
    \\
    & = &
    \frac{1}{|\Z_2|} (|\Z_2|Z_{T^3}([X/\Z_2]) + 2Z_{T^3}(X)), 
    \\
    &= &Z_{T^3}([X/\Z_2]) + Z_{T^3}(X),
\end{eqnarray*}
which matches the result above.

\subsection{$\Gamma=(\Z_4\xrightarrow{0}\Z_2)$}

Consider the crossed module $(\Z_4\xrightarrow{0}\Z_2)$ with action $\alpha(z)(a)=a^3$,
where $G = {\mathbb Z}_2 = \langle z \rangle$, $K = {\mathbb Z}_4 = \langle a \rangle$. 
In this case, in the notation of the decomposition conjecture~(\ref{eq:basic-conj}), $G = {\mathbb Z}_2$,
$K = {\mathbb Z}_4$, and $G$ leaves invariant two of the irreducible representations of $K$
but exchanges the other two.  The decomposition conjecture~(\ref{eq:basic-conj}) then predicts
\begin{eqnarray}
    {\rm QFT}\left( [X/\Gamma] \right)
    & = &
    {\rm QFT}\left( \left[ \frac{ X \times \hat{A}}{G} \right] \right),
    \\
    & = &
    {\rm QFT}\left( [X/{\mathbb Z}_2] \, \coprod \, [X/{\mathbb Z}_2] \, \coprod \,
    \left[ \frac{ X \times {\mathbb Z}_2}{ {\mathbb Z}_2 } \right] \right),
    \\
    & = &
    {\rm QFT}\left( [X/{\mathbb Z}_2] \, \coprod \, [X/{\mathbb Z}_2] \, \coprod \,
    X  \right).
\end{eqnarray}

Next, we compare to physics.
The partition function is
\begin{eqnarray*}
    Z_{T^3}(\Z_4\xrightarrow{0}\Z_2) 
    & = &
    \frac{1}{|\Z_4|^2|\Z_2|} \bigl(|\Z_4|^3Z(1,1,1) 
    \nonumber \\
    & & \hspace*{1in}
    + |\Z_4|^2|\Z_2|\left[Z(1,1,z)+Z(1,z,z)+\text{perms.}+Z(z,z,z)\right]\bigr),
    \\
    & = &
    Z(1,1,1)+(Z(1,1,1)+Z(1,1,z)+Z(1,z,z)+\text{perms.}+Z(z,z,z)),
    \\
    & = &
    Z_{T^3}(X)+|\Z_2|Z_{T^3}([X/\Z_2]),
\end{eqnarray*}
in agreement with the decomposition prediction computed above.

\subsection{$\Gamma=(\Z_3\times\Z_3\xrightarrow{0}\Z_2)$}

Consider the crossed module
$(\Z_3\times\Z_3\xrightarrow{0}\Z_2)$ with action $\alpha(z)(a)=a^2$, $\alpha(z)(b)=b^2$,
where $G = {\mathbb Z}_2 = \langle z \rangle$, $K = {\mathbb Z}_3 \times {\mathbb Z}_3  = \langle a \rangle \times \langle b \rangle$. 
In terms of the decomposition conjecture~(\ref{eq:basic-conj}), 
%$G = {\mathbb Z}_2$,
%$K = {\mathbb Z}_3 \times {\mathbb Z}_3$, and 
the action of $G$ on $\hat{K}$ leaves one representation
invariant (the trivial representation), while interchanging the other eight irreducible representations
in pairs.  The decomposition conjecture then predicts
\begin{eqnarray}
    {\rm QFT}\left( [X/\Gamma] \right)
    & = &
    {\rm QFT}\left( \left[ \frac{X \times \hat{K}}{G} \right]\right),
    \\
    & = &
    {\rm QFT}\left( [X/{\mathbb Z}_2] \, \coprod \, \left[ \frac{ X \times ( {\mathbb Z}_2)^4 }{{\mathbb Z}_2} \right] \right),
    \\
    & = &
    {\rm QFT}\left(  [X/{\mathbb Z}_2] \, \coprod \, \coprod_4 X \right).
\end{eqnarray}
The first summand is associated to the trivial representation of $K = {\mathbb Z}_3 \times {\mathbb Z}_3$,
and the other four correspond to the four orbits of the ${\mathbb Z}_2$ action on $\hat{K}$.

The partition function is
\begin{eqnarray*}
    Z_{T^3}((\Z_3\times\Z_3\xrightarrow{0}\Z_2))
    & = &
    \frac{1}{|\Z_3\times\Z_3|^2|\Z_2|} \bigl( |\Z_3\times \Z_3|^3Z(1,1,1)
    \nonumber \\
    & & \hspace*{1.0in}
    +|\Z_3\times\Z_3|^2 \bigl[ Z(1,1,z)
    %\nonumber \\
    %& & \hspace*{1.5in}
    +Z(1,z,z)+\text{perms.}+Z(z,z,z)\bigr]\bigr),
    \\
    & =& 
    \frac{1}{|\Z_2|}\left(8Z(1,1,1)+Z(1,1,1)+Z(1,z,z)+\text{perms.}+Z(z,z,z)\right),
    \\
    & = &
    Z_{T^3}(X/\Z_2) + 4Z_{T^3}(X),
\end{eqnarray*}
matching the prediction of the decomposition conjecture.

\subsection{$\Gamma=(\Z_3\xrightarrow{0}\Z_2\times\Z_2)$}

Let $\Gamma=(\Z_3\xrightarrow{0}\Z_2\times\Z_2)$ with $\alpha(a)(z)=z^2$ and $\alpha(b)(z)=z$, where
$G = {\mathbb Z}_2 \times {\mathbb Z}_2 = \langle a \rangle \times \langle b \rangle$ and
$K = {\mathbb Z}_3 = \langle z \rangle$. That is, for the action $\alpha:\Z_2\times\Z_2\to\text{Aut}(\Z_3)=\Z_2$ one has that $\text{ker}(\alpha)=\langle b \rangle$ and $a\in\Z_2\times\Z_2$ injects into $\text{Aut}(\Z_3)$. 

Let us apply the decomposition prediction~(\ref{eq:basic-conj}).  Here, $b \in G$ acts trivially on
$\hat{K}$, whereas $a \in G$ leaves one irreducible representation of $K$ invariant (the trivial representation) and exchanges the other two.  Furthermore, since $\delta = 0$, $\beta = 0$, so there is no discrete torsion on any summand.  Then,
from Equation~(\ref{eq:basic-conj}), decomposition predicts
\begin{eqnarray}
    {\rm QFT}\left( [X/\Gamma] \right) 
    & = & {\rm QFT}\left( \left[ \frac{X \times \hat{K} }{G} \right] \right),
    \\
    & = &
    {\rm QFT}\left( [X/G] \, \coprod \, \left[ \frac{X \times {\mathbb Z}_2 }{ {\mathbb Z}_2 \times {\mathbb Z}_2 } \right] \right),
    \\
    & = &
    {\rm QFT}\left( [X/ {\mathbb Z}_2 \times {\mathbb Z}_2] \, \coprod \, [X/{\mathbb Z}_2] \right).
\end{eqnarray}

Next, we compute the partition function on $T^3$.
In the present case, the group elements $1,b\in\klein$ fix all elements of $\Z_3$, so that the twisted sectors that only involve $1,b$ will have a $|\Z_3|^3$-fold multiplicity. On the other hand, $a,ab\in\klein$ do not fix any nontrivial elements in $K = \Z_3$.  Assembling the pieces, one finds that
the partition function is
\begin{eqnarray*}
    Z_{T^3}(\Z_3\xrightarrow{0}\Z_2\times\Z_2) 
    & = &
    \frac{1}{|\Z_3|^2|\klein|} ( |\Z_3|^3 (\ \text{sectors involving \underline{only} $\{1,b\}$}) 
    \nonumber \\
    & & \hspace*{1.25in}
    + |\Z_3|^2(\ \text{sectors involving $a$})),
    \\
    & = &
    \frac{1}{|\klein|} (2 Z(\text{\underline{only}}\ \{1,b\})+ Z(\text{\underline{only}}\ \{1,b\}) + Z(\{a\})),
    \\
    & = &
    \frac{2}{|\klein|} Z(\text{\underline{only}}\ \{1,b\}) + \frac{1}{|\klein|}Z(\{1,a,b,ab\}),
    \\
    & = &
    Z_{T^3}([X/\Z_2]) + Z_{T^3}([X/\klein]),
\end{eqnarray*}
matching the decomposition prediction above.

Note that, unlike previous examples, here we do not get a copy of $Z(X)$. This is because there is a nontrivial group element $b\in\klein$ that acts trivially on $\Z_3$, so that its twisted sectors have the same multiplicity as $Z(1,1,1)$. This seems to suggest that we when gauging a 2-group, one of the components will be $Z([X/\text{ker}\ \alpha])$.

\subsection{$\Gamma=(\Z_3\times\Z_3\xrightarrow{0}\klein\times\Z_2)$}

We take $\Gamma=(\Z_3\times\Z_3\xrightarrow{0}\klein\times\Z_2)$ with $\alpha(a)(x,y)=(x,y)$, $\alpha(b)(x,y)=(x^2,y)$ and $\alpha(c)(x,y)=(x,y^2)$,
where
\begin{equation}
    G \: = \: {\mathbb Z}_2 \times {\mathbb Z}_2 \times {\mathbb Z}_2 \: = \:
    \langle a \rangle \times \langle b \rangle \times \langle c \rangle,
    \: \: \:
    K \: = \: {\mathbb Z}_3 \times {\mathbb Z}_3 \: = \: \langle x \rangle \times \langle y \rangle.
\end{equation}

We begin by computing the decomposition prediction~(\ref{eq:basic-conj}).
To that end, $a \in G$ leaves $\hat{K}$ invariant, $b \in G$ interchanges six of the nine isomorphism
classes in $\hat{K}$, in two sets of three, and $c \in G$ similarly interchanges another six of the
nine elements of $\hat{K}$.  One of the elements of $\hat{K}$ (corresponding to the trivial representation) is invariant under all of $G$.  Four of elements of $\hat{K}$ are acted upon by two ${\mathbb Z}_2$, two are exchanged by only one ${\mathbb Z}_2$, and another two are exchanged by only another.  Also, since $\delta = 0$, we have $\beta = 0$, hence there is no discrete
torsion on any universe.  Assembling these pieces, we have
\begin{eqnarray}
    {\rm QFT}\left( [X/\Gamma] \right) & = &
    {\rm QFT}\left( \left[ \frac{X \times \hat{K} }{G} \right] \right),
    \\
    & = & 
    {\rm QFT}\biggl(  [X/G] \, \coprod \, \left[ \frac{X \times {\mathbb Z}_2 \times {\mathbb Z}_2 }{ {\mathbb Z}_2 \times {\mathbb Z}_2 \times {\mathbb Z}_2} \right] 
    \nonumber \\
    & & \hspace*{0.5in}
    \, \coprod \,
    \left[ \frac{X \times {\mathbb Z}_2 }{ {\mathbb Z}_2 \times {\mathbb Z}_2 \times {\mathbb Z}_2} \right] 
    \, \coprod \,  \left[ \frac{X \times {\mathbb Z}_2 }{ {\mathbb Z}_2 \times {\mathbb Z}_2 \times {\mathbb Z}_2} \right] 
    \biggr),
    \\
    & = &
    {\rm QFT}\left( [X/G] \, \coprod \, [X/{\mathbb Z}_2] \, \coprod \,
    [X/{\mathbb Z}_2 \times {\mathbb Z}_2] \, \coprod \,
    [X/{\mathbb Z}_2 \times {\mathbb Z}_2] \right).
\end{eqnarray}
The first summand corresponds to the trivial (and $G$-invariant) representation in $\hat{K}$, the next one to the group of four acted upon nontrivially by two ${\mathbb Z}_2$s, and the last two summands correspond to elements exchanged by only one of the ${\mathbb Z}_2$s.

Note that $\text{ker}\ \alpha$ is generated by $a$, and that both $b$ and $c$ have $|\Z_3|$ fixed points, but $bc$ has 1 fixed point. Then
\begin{eqnarray*}
    Z\left([X/\Gamma]\right) 
    & = &
    \frac{1}{|\Z_3\times\Z_3|^2|\klein\times\Z_2|} \bigl(|\Z_3\times\Z_3|^3 Z(\{1,a\}) \\
    & & \hspace*{1.75in} + |\Z_3\times\Z_3|^2|\Z_3| (Z(1,1,b)+Z(1,1,c))\\
    & & \hspace*{1.75in}  +|\Z_3\times \Z_3|^2 Z(1,1,bc)+\cdots \bigr), \\
    & = &
    Z_{T^3}(X/\klein\times\Z_2])  \\
    & & + \frac{1}{|\Z_2|^3}(8Z(\{1,a\})%\\
    %& & 
    + 2Z(1,1,b)+2Z(1,1,c)+\cdots), \\
    & = & Z_{T^3}([X/\klein\times\Z_2])+2\cdot Z_{T^3}([X/\klein])+Z([X/\Z_2]),
\end{eqnarray*}
matching the prediction above.

\subsection{$G=\Z_2$, $K=\Z_4$ with nontrivial action and extension}
We consider the $2$-group $\Gamma$ arising from the fibration
\begin{equation*}
    B^2\Z_4\longrightarrow B\Gamma\longrightarrow B\Z_2
\end{equation*}
such that the action $\alpha:\Z_2\to\text{Aut}(\Z_4)$ is $^za=a^3$, and the extension class $[\beta_{\Gamma}]\in H^3(\Z_2,\Z_4)=\Z_2$ is the nontrivial one. In this case
\begin{equation}
    G\: = \: \Z_2\: =\: \langle z\rangle, \: \: \: K \: = \: \Z_4 \: = \: \langle a\rangle
\end{equation}

We compute the decomposition prediction~(\ref{eq:basic-conj}). We have that $\hat{K}=\Z_4$, and that the generator $z\in G$ interchanges the generator $\hat{k}\in\hat{K}$ with $\hat{k}^3$ while leaving $0,\hat{k}^2\in\hat{K}$ invariant. Thus, we have three distinct orbits: $\{0\}$, $\{\hat{k^2}\}$, and $\{\hat{k},\hat{k}^3\}$. The discrete torsion contributions for each universe in the prediction come from cocycles that are expressed as $\rho\circ\beta$ for $\beta$ the extension class. Since we are working with normalized cocycles, only $\beta(z,z,z)$ can have a nontrivial value. However, the discrete torsion-like associated factor for this triple is trivial, as
\begin{equation}\label{eq:dtlikephasez2z4}
    \epsilon_{\beta_{\Gamma}}(z,z,z)= \frac{\beta_{\Gamma}(z,z,z)}{\beta_{\Gamma}(z,z,z)}\frac{\beta_{\Gamma}(z,z,z)}{\beta_{\Gamma}(z,z,z)}\frac{\beta_{\Gamma}(z,z,z)}{\beta_{\Gamma}(z,z,z)}=1
\end{equation}
vanishes identically. This means that, although the extension class is nontrivial, the discrete torsion phases entering the decomposition are all trivial. Thus, the conjecture reads
\begin{eqnarray}
    {\rm QFT}\left( [X/\Gamma] \right) & = &
    {\rm QFT}\left( \left[ \frac{X \times \hat{K} }{G} \right] \right),
    \\
    & = & 
    {\rm QFT}\biggl(  [X/G] \, \coprod \, [ X/G ]_{\omega} 
    \coprod \left[\frac{X\times\Z_2}{\Z_2}\right]\biggr),
    \\
    & = &
   {\rm QFT}\biggl(  [X/G] \, \coprod \, [ X/G ] 
    \coprod [X]\biggr).\label{eq:z2z4conj}
\end{eqnarray}

We now compare this to a computation at the level of partition functions. This exemplifies a computation using only the invariant data that defines a $2$-group, without making use of an explicit crossed module presentation. As observed in Equation (\ref{eq:dtlikephasez2z4}), all the associated discrete torsion-like factors are trivial. This means that all triples $(g,h,k)\in\Z_2\times\Z_2\times\Z_2$ are admissible, according to Equation~(\ref{eq:Z:invt}). The $3$-torus partition function is then
\begin{eqnarray*}
    Z_{T^3}([X/\Gamma]) &=& \frac{1}{|\Z_4|^2 |\Z_2|}\sideset{}{'}\sum_{k_1,k_2,k_3\in \Z_4}\sideset{}{'}\sum_{g,h,k\in \Z_2} Z(g,h,k;k_1,k_2,k_3),
    \\
   &=& \frac{1}{|\Z_4|^2 |\Z_2|} (|\Z_4|^3Z(1,1,1)+|\Z_4|^2|\Z_2|(Z(1,1,z)+Z(1,z,z)+Z(z,z,z)+\text{perms.})),
   \\
   &=& Z_{T^3}(X)+Z_{T^3}([X/\Z_2])+Z_{T^3}([X/\Z_2]),
\end{eqnarray*}
matching what is predicted by the conjecture (\ref{eq:z2z4conj}). This is an example where even though a nontrivial extension class is used, the result is identical to what would be obtained in the case with trivial extension class.

\section{Conclusions}

In this paper we have analyzed 2-group orbifolds $[X/\Gamma]$ of low-energy effective theories in three dimensions, generalizing the results of \cite{Pantev:2022kpl}, which focused on 2-groups $\Gamma$ which were analogues of central extensions (meaning, trivial $\alpha$).  On the assumption that the gauged one-form symmetry acts trivially, we have described partition function computations, and also discussed the form of decomposition in these theories.  We have found a more general pattern of decomposition in three-dimensional theories than appeared in previous three-dimensional examples.

\section{Acknowledgements}

The authors would like to thank D.~Robbins and T.~Pantev for useful comments.
A.P.L.~would like to thank D.~Pavlov, and U.~Schreiber for useful suggestions on cohomology computations. E.S.~was partially supported by NSF grant PHY-2014086.

\appendix

\section{Background on 2-groups}
\label{app:background}

\subsection{2-groups as crossed modules}\label{ssec:xmod}

In general terms, the concept of a 2-group may be regarded as a categorification of a group. In the literature, it has appeared under various equivalent guises. In the present article, however, the presentations that we will mainly use are those of crossed modules, for computations, and of one-object 2-groupoids. The first approach is reviewed in this section, whereas the second approach is reviewed in Section \ref{ssec:2grpds}. We also briefly review the presentation as Lie 2-groups in \ref{ssec:2grpact}, though only to clarify what a 2-group action on a space actually means. 

While crossed modules technically represent \textit{strict} 2-groups, it is known, as explained for example in \cite{elgueta2014permutation}, that every (weak) 2-group is isomorphic to a strict 2-group. This presentation has the advantage that it reduces the computations involving 2-groups to computations involving traditional groups and homomorphisms between them.

We recall the definition of a crossed module, due to Whitehead \cite{whitehead49}. A \textit{crossed module} $\Gamma=(\Gamma_1\xrightarrow{\delta} \Gamma_0)$ consists of a pair of groups $\Gamma_0, \Gamma_1$, endowed with a homomorphism $\delta:\Gamma_1\to \Gamma_0$ called the \textit{boundary} map, and an action $\overline{\alpha}:\Gamma_0\to\text{Aut}(\Gamma_1)$ of $\Gamma_0$ on $\Gamma_1$. For simplicity, we denote the action of $g\in\Gamma_0$ on $h\in\Gamma_1$ as $\overline{\alpha}(g)(h)=\tensor[^g]{h}{}$. These homomorphisms are required to satisfy $\Gamma_0$-equivariance
\begin{equation}
    \delta(\tensor[^g]{h}{}) = g\delta(h)g^{-1}
\end{equation}
and the \textit{Pfeiffer identity}
\begin{equation}
      % \tensor[^{\delta(h)}]{h'}{} = h h' h^{-1}  \: \: \:
        {}^{ \delta(h) }h' \: = \: h h' h^{-1}
   \end{equation}

In general, there may exist several different choices of $\Gamma_0$ and $\Gamma_1$ that represent the same crossed module. Up to isomorphism, any crossed module $\Gamma=(\Gamma_1\xrightarrow{\delta}\Gamma_0)$ is uniquely specified by the following data \cite{Baez:2005sn}:
\begin{itemize}
    \item The cokernel $\cokerd$.
    \item The kernel $\kerd$, which is abelian.
    \item The Postnikov extension class $[\beta]\in H^3(\cokerd,\kerd)$, which specifies a short exact sequence
    \begin{equation}
    1\to\kerd\hookrightarrow\Gamma_1\to\Gamma_0\twoheadrightarrow\cokerd\to 1
\end{equation}
    \item An action $\alpha:\cokerd\to\aut (\kerd)$.
\end{itemize}

An obvious but important fact is that the inclusion $\kerd\hookrightarrow \Gamma_1$ factors through the center $Z(\Gamma_1)$ of $\Gamma_1$ \cite{weibel1995introduction} as
\begin{equation}
    \kerd\hookrightarrow Z(\Gamma_1)\hookrightarrow \Gamma_1.
\end{equation}

The fact that crossed modules are specified by the four ingredients listed above establishes a \textit{weak equivalence} amongst them. To see how this works, we first need to define the morphisms of crossed modules. 

For crossed modules $\xmod$ and $\Gamma'=(\Gamma_1'\xrightarrow{\delta'}\Gamma_0')$, a homomorphism of crossed modules $(f,\phi):\Gamma\to\Gamma'$ consists on a pair of group homomorphisms $f:\Gamma_1\to\Gamma_1'$ and $\phi:\Gamma_0\to\Gamma_0'$ subject to the conditions that $\delta'\circ f= \phi\circ\delta$ and that $f(\tensor[^g]{t}{})=\tensor[^{\phi(g)}]{f(t)}{}$ for $g\in\Gamma_0$ and $t\in\Gamma_1$. Furthermore, we say that a crossed module homomorphism $(f,\phi):\Gamma\to\Gamma'$ is a \textit{weak equivalence} if the induced homomorphisms $\kerd\to\text{ker}(\delta')$ and $\cokerd\to\text{coker}(\delta')$ are isomorphisms. Topologically, this means that $\Gamma$ and $\Gamma'$ represent the same topological space up to weak equivalence of topological spaces.

\subsection{2-group actions}\label{ssec:2grpact}
Since the concept of action of a 2-group on a smooth manifold will play an important role here, we quickly recall the relevant definitions. 

Actions of 2-groups on smooth manifolds are more clearly seen as actions of Lie 2-groups on Lie groupoids. Both of these concepts are in turn easily understood using the concept of an internal category, which we recall now \cite{cthandbook}. Intuitively, the definition captures the properties that define a category so that it can be regarded as a category inside another category. Here, we assume that we can talk about the points of the objects, which will be the case of interest. In some ambient category $\mathcal{C}$, an internal category $C$ consists on a pair of objects $C_0, C_1\in\text{ob}(\mathcal{C})$. The interpretation $C_0, C_1$ is that of object of objects, and object of morphisms. These are endowed with morphisms $s,t:C_1\to C_0$, $i:C_0\to C_1$, and $\circ:C_1\times_{C_0}C_1\to C_1$ where $C_1\times_{C_0}C_1 = \{(f,g)\in C_1\times C_1 \ \text{s.t.} \ t(f)=s(g)\}$. The morphisms $s,t,i$ are to be viewed as giving the source and target objects of morphisms in $C$, and the identity morphisms, respectively. The morphism $\circ$ describe composition of morphisms where the target object of one is the source object of the other.

We now recall the definition of a functor of internal categories. Given two internal categories $C$ and $D$, a functor of internal categories $F:C\to D$ consists of a pair of morphisms $F_0:C_0\to D_0$ and $F_1:C_1\to D_1$ such that $s\circ F_1 = F_0\circ s$, $t\circ F_1= F_0\circ t$, $F_1 \circ i = i\circ F_0$, and $F_1\circ c=c\circ(F_1\times_{F_0} F_1)$.

Using this definition one can define Lie groupoids and Lie 2-groups. See e.g.~\cite{mackenzie,Baez:2002jn} for more information. A Lie groupoid $C$ is a category internal to the category of manifolds. This means that $C_0, C_1$ are manifolds, and the maps involved in the definition are all smooth. Similarly, a Lie 2-group $G$ is a category internal to the category of Lie groups, meaning the objects $G_0, G_1$ are Lie groups, and the involved morphisms are smooth Lie homomorphisms. Note that Lie 2-groups may be regarded as special cases of Lie groupoids.

We now recall the definition of a Lie 2-group action on a Lie groupoid. Let $G$ be a Lie 2-group and $X$ a Lie groupoid. An action of $G$ on $X$ consists of Lie group actions $\mu_0:G_0\times X_0\to X_0$ and $\mu_1:G_1\times X_1\to X_1$, such that they form a Lie groupoid functor from the Lie groupoid $(G_1\times X_1 \to G_0\times X_0)$ to $X$.

This machinery now allows us to describe the action of a 2-group $\Gamma$ on the target space $X$. First, we need to present $X$ as a Lie groupoid. There are several Lie groupoids that one can construct from $X$. Here, we choose to present $X$ as the Lie groupoid with $X_0=X$, $X_1=X$, and $s=t=i=\text{Id}_X:X\to X$ the smooth identity map. We choose this particular presentation because the orbifold theory describes maps $f:T^2\to X$, and we do not want to broaden the allowed maps when presenting $T^2$ and $X$ as Lie groupoids. This, however, is an interesting generalization that we would like to consider in the future.

Now we present our crossed module $\Gamma=(\Gamma_1\to \Gamma_0)$ of finite groups as a Lie 2-group $\hat{\Gamma}$. We take $\hat{\Gamma}_0= \Gamma_0$ and $\hat{\Gamma}_1 = \Gamma_1\rtimes \Gamma_0$, where each is given the smooth structure of discrete groups, so as to be considered Lie groups. The semidirect product $\hat{\Gamma}_1$ is defined by the group law 
\begin{equation}
(h,g)\cdot (h',g') = (h ^g h',gg')
\end{equation}
for $^g h'$ the action of $g$ on $\Gamma_1$ through the homomorphism $\alpha: \Gamma_0\to \text{Aut}(\Gamma_1)$. We take $s=\pi_{\Gamma_0}$ the projection $\pi_{\Gamma_0}:\Gamma_1\rtimes \Gamma_0\to \Gamma_0$, $t(h,g)=\delta(h)g$, and $i(g)=(1,g)$.

It is now clear what it means for $\hat{\Gamma}$ to act on $X$. An action of a Lie 2-group $\hat{\Gamma}$ on a smooth manifold $X$ is a Lie 2-group action on the canonical Lie groupoid associated with $X$. It consists on two compatible group actions $\mu_1: (\Gamma_1\rtimes \Gamma_0)\times X\to X$ and $\mu_0: \Gamma_0\times X\to X$, such that $s_X\circ \mu_1 = \mu_0 \circ (s_{\hat{\Gamma}}\times s_X)$ and $t_X\circ \mu_1 = \mu_0 \circ (t_{\hat{\Gamma}}\times t_X)$, meaning that the diagram
\begin{center}
\begin{tikzcd}
(\Gamma_1\rtimes \Gamma_0)\times X \arrow[rr, "\mu_1"] \arrow[dd, "s_{\hat{\Gamma}}\times s_X"', bend right] \arrow[dd, "t_{\hat{\Gamma}}\times t_X", bend left, shift left] &  & X \arrow[dd, "s_X"', bend right] \arrow[dd, "t_X", bend left, shift left] \\
                                                                                                                                                                             &  &                                                                           \\
\Gamma_0\times X \arrow[rr, "\mu_0"]                                                                                                                                         &  & X                                                                        
\end{tikzcd}
\end{center}
commutes.

One of the consequences of this definition is that $\Gamma_1$ automatically acts trivially on $X$. Note that $s_X\circ \mu_1 ((h,1),m) = (h,1)\cdot m$. By the required commutativity this is
\begin{equation}
    \mu_0 \circ (s_{\hat{\Gamma}}\times s_X) ((h,1),m) = \mu_0(1,m)=m
\end{equation}
so that $\mu_1$ is a trivial action. Of course, not only does $\Gamma_1$ act trivially but also its image $\text{Im}(\delta)\subset\Gamma_0$. All in all, this means that the only action that potentially is nontrivial is that of $\cokerd= G$ on $X$, and in particular $\kerd=K$ acts trivially.  

To summarize, the action of a 2-group on a space $X$ factors through an action of
$G = {\rm coker}\, \delta$, such that the image of $\delta$ acts trivially.

\subsection{2-groups as one-object 2-groupoids}\label{ssec:2grpds}

We end this section by reviewing how 2-groups may also be regarded as 2-groupoids with one object. This definition is not used in the present paper for explicit computations, but it is used in the derivation of the partition function. As described in Section \ref{ssec:2grpact}, a crossed module $\Gamma=(\Gamma_1\xrightarrow{\delta}\Gamma_0)$ may be regarded as a category whose space of objects and morphisms are (Lie) groups. A closely-related presentation is given in terms of a 2-category. Broadly speaking, a 2-category is a generalization of a category where one has objects, morphisms, and morphisms between morphisms \cite{kelly2006review}. One usually refers to objects, 1-morphisms, and 2-morphisms. In particular, a 2-groupoid is a 2-category where 1- and 2-morphisms are invertible. The precise meaning for invertibility depends on whether we consider weak or strict 2-categories. Here we mainly work with strict 2-categories, where 1-morphisms are exactly invertible and not only up to a 2-morphisms.

Hence, a different way to present a 2-group given by a crossed module $\Gamma=(\Gamma_1\xrightarrow{\delta}\Gamma_0)$ is as a 2-groupoid with a single object, whose 1-morphisms are $\Gamma_0$, and whose 2-morphisms are $\Gamma_1\rtimes \Gamma_0$. This definition is identical to the presentation as Lie 2-groups, but shifted one categorical degree up. This presentation is helpful in particular because it allows for a clean realization as a simplicial set thanks to the Duskin nerve \cite{duskin1975simplicial}, so that its connection to cohomology and triangulation of topological spaces is straightforward.

It is known that there exist several conventions for representing crossed modules as 2-groupoids with a single object. Throughout this paper, we use the \textbf{LB} convention. See \cite{rob08} for more details.

\end{document}